\documentclass[AMA,STIX1COL]{WileyNJD-v2}

\usepackage{psfrag}
\usepackage{amsmath}
\usepackage{amsthm}
\usepackage{subfigure}
\usepackage{mathtools}
\usepackage{epsfig}
\usepackage{amsfonts}
\usepackage{dsfont}
\usepackage{siunitx}

\newtheorem{defin}{{Definition}} 

\articletype{Reseach Article}%

\begin{document}
	
	\title{A family of virtual contraction based controllers for tracking  of  flexible-joints port-Hamiltonian robots: theory and experiments\protect\thanks{Partial results were presented in the IFAC Workshop on Lagrangian and Hamiltonian Methods in Nonlinear Control 2018.}}
	
	\author[1,2]{Rodolfo Reyes-B\'aez*}
	
	\author[1,2]{Arjan van der Schaft}
	
	\author[1,3]{Bayu Jayawardhana}
	
	\author[3]{Le Pan}
	
	\authormark{R. Reyes-B\'aez \textsc{et al}}

	\address[1]{\orgdiv{Jan C. Willems Center for Systems and Control}, \orgname{University of Groningen}, \orgaddress{\state{Groningen}, \country{The Netherlands}}}
	
	\address[2]{\orgdiv{Bernoulli Institute for Mathematics, Computer Science and Artificial Intelligence}, \orgname{University of Groningen}, \orgaddress{\state{Groningen}, \country{The Netherlands}}}
	
	\address[3]{\orgdiv{Engineering and Technology Institute Groningen (ENTEG)}, \orgname{University of Groningen}, \orgaddress{\state{Groningen}, \country{The Netherlands}}}
	
	\corres{*Rodolfo Reyes-B\'aez, Faculty of Science and Engineering, University of Groningen,  Groningen, The Netherlands. \email{r.reyes-baez@ieee.org}}
	

	
	\abstract[Summary]{
		In this work we present a constructive method to design  a family  of virtual contraction based controllers that solve the standard trajectory tracking problem of  flexible-joint robots (FJRs) in the port-Hamiltonian (pH) framework. The proposed design  method, called virtual contraction based  control (v-CBC), combines the concepts of virtual control systems and  contraction analysis. It is shown that under potential energy matching conditions, the closed-loop virtual system is contractive and   exponential convergence to a predefined trajectory is guaranteed. Moreover, the closed-loop virtual system   exhibits     properties such as structure preservation, differential passivity and the existence of  (incrementally) passive  maps. \\
	}
	
	\keywords{Flexible-joints robots, tracking control, port-Hamiltonian systems, contraction, virtual control systems}
	
	
	\maketitle
	
	
	\section{Introduction}
	Control  problems in rigid robots have been widely studied  in the literature due to they are  instrumental in modern manufacturing systems. However, as pointed out  in Tomei  \cite{tomei1995tracking} the elasticity  in the joints often can not be neglected for accurate position tracking.  For every joint that is actuated by a motor, we basically need two degrees of freedom  instead of one. Such FJRs are therefore \emph{underactuated} mechanical systems. In the work of Spong\cite{spongflexible} two state feedback control laws based,  respectively, on  feedback linearization and singular perturbation theory are presented for a simplified FJRs model. Similarly, in Canudas\cite{canudas} a dynamic feedback controller for a more detailed model is presented. In Lor\'ia \cite{loria1995tracking} a computed-torque controller for FJRs is designed, which does not need \emph{jerk} measurements.  In Ortega\cite{ortega-regulacion} and Brogliato\cite{brogliato1995global}   passivity-based control (PBC) schemes are proposed. The first one is an observer-based controller which requires only motor position measurements. In the latter one, a PBC controller is designed and compared with backstepping and decoupling techniques. For further details on PBC of FJRs we refer to Ortega et al. \cite{ortega2013passivity} and references therein. In Astolfi \cite{astolfi2003immersion}, a global tracking controller based on the immersion and invariance (I\&I) method is introduced.\\
	From a  practical point of view, in Albu-Sch\"affer \cite{albu2007unified}, a torque feedback is embedded into the passivity-based control approach, leading to a full state feedback controller, where   acceleration and jerk measurements are not required. In the  recent work of  \'Avila-Becerril \cite{Sofia},  a dynamic controller is designed which solves the global position tracking problem  of FJRs based only on measurements of link and joint positions. In the work of \cite{pan2017adaptive} an adaptive-filtered backstepping design is experimentally evaluated in a single flexible-joint prototype. All of these control methods are designed for FJRs modeled as  second order Euler-Lagrange (EL) systems. Most of these schemes are based on the selection of a suitable storage function that together with the dissipativity of the closed-loop system,  ensures the convergence of the  state trajectories to the desired solution. 
	
	As an alternative to the EL formalism, the pH framework has been introduced in van der Schaft\cite{vanderschaft1995}. The main characteristics of the pH framework are the existence of a Dirac structure (connects geometry with analysis),  port-based network modeling and  the \emph{clear physical energy interpretation}. For the latter part, the energy function can directly be used to show the dissipativity   of the systems.  Some  set-point  controllers have been proposed for FJRs modeled as pH systems. For instance in Borja \cite{Borja2014flexible} the controller for FJRs modeled as EL systems in Ortega\cite{ortega2013passivity} is adapted and interpreted in terms of  the Control by Interconnection technique\footnote{We refer interested readers on  CbI to   \cite{castanos}.} (CbI). In Zhang\cite{chinoflexiblejoint}, they propose an Interconnection and Damping Assignment  PBC (IDA-PBC\footnote{For IDA-PBC technique see also \cite{escobar}.}) scheme, where the controller is designed with respect to the pH representation of the  EL-model  in  Albu-Sch\"affer\cite{albu2007unified}. 
	
	 For the tracking control case of FJRs in the pH framework, to the best of our knowledge, the only results available in the literature are the singular-perturbation approach in Jard\'on-Kojakhmetov\cite{hilde} and our preliminary work Reyes-B\'aez\cite{reyes2017virtual}. 
	
	In the present  work we propose a setting that extends our previous  results in  Reyes-B\'aez\cite{reyesIFAC2017} and  Reyes-B\'aez\cite{RodoAutomatica2017}  on  v-CBC of fully-actuated mechanical  systems  to solve the tracking problem of FJRs modeled as pH systems. This  method  relies on the \emph{contraction} properties of the so-called virtual system, see the works\cite{forni, pavlov2017convergent,slotinecontraction,sontag2010contractive,wang}.   Roughly speaking, the  method\footnote{The  use of virtual systems for control design  was already considered in \cite{jouffroy} and \cite{manchester2015unifying}.} consists in designing a control law for a  virtual system associated to the \emph{original} FJR, such that the closed-loop virtual system is contractive  and a predefined reference trajectory is exponentially stable. Finally, this control scheme is applied to the original FJR. It follows that the reference trajectory  of the virtual system and the original state converge to each other.
	
	The paper is organized as follows: In Section 2, the theoretical  preliminaries  on virtual contraction based control (v-CBC) and key properties of mechanical systems in the pH framework are presented. Section 3 presents the  pH model of FJRs, together with the statement of the trajectory tracking problem and its solution. The main result on the construction of a  family v-CBC schemes for FJRs  are presented in Section 4. In Section 5, the performance of two v-CBC tracking controller is evaluated experimentally on a two-degrees of freedom FJR. Finally, in Section 6 conclusions and future research are stated.

	\section{Preliminaries}\label{sII}
	\subsection{Contraction analysis and differential passivity}
	In this section, the differential approach to incremental stability\cite{angeli2002lyapunov} by means of contraction analysis is summarized. Sufficient conditions   in terms of the frameworks of the  differential Lyapunov theory  \cite{forni} and of the matrix measure\cite{sontag2010contractive} are given. These ideas are later extended to  systems having inputs and outputs  with the notion of differential passivity\cite{arjan2013differentialpassivity}, and to virtual control systems\cite{wang,jouffroy}. For a self-contained and detailed introduction to these topics see also \cite{ReyesBaez-PhDthesis}.\\
	
	Let $\mathcal{X}$ be an $N$-dimensional state space manifold    with  local  coordinates $x=(x_1,\dots,x_N)$ and tangent bundle $T\mathcal{X}$. Let   $\mathcal{U}\subset\mathds{R}^n$ and $\mathcal{Y}\subset\mathds{R}^n$ be the input and output spaces, respectively. Consider the nonlinear control system $\Sigma_u$,  affine  in the input $u$,   given by
	\begin{equation}
	\Sigma_u:\left\{ \begin{array}{llc}
	\dot{{x}}={f}(x,t)+\sum_{i=1}^{n}{g}_i(x,t){u}_i,\\
	y=h(x,t),
	\end{array}
	\right.	
	\label{eq:controlsystem}
	\end{equation}
	where  $x\in\mathcal{X}$, ${u}\in\mathcal{U}$  and $y\in\mathcal{Y}$. The time varying vector fields ${f}:\mathcal{X}\times\mathds{R}_{\geq 0}\rightarrow T\mathcal{X}$,  ${g}_i:\mathcal{X}\times\mathds{R}_{\geq 0}\rightarrow T\mathcal{X}$ for $i\in\{1,\dots,n\}$ and the output function $h:\mathcal{X}\times\mathds{R}_{\geq 0}\rightarrow \mathcal{Y}$  are  assumed to be smooth.  System $\Sigma_u$ in closed-loop with the state feedback $u=\gamma(x,t)$ defines the system $\Sigma$ given by
	\begin{equation}
	\begin{split}
	\Sigma:\left\{ \begin{array}{lcc}
	\dot{{x}}={F}({x},t)={f}(x,t)+\sum_{i=1}^{n}{g}_i(x,t)\gamma_i(x,t), \\
	{y}={h}(x,t).		
	\end{array}
	\right.
	\end{split}
	\label{eq:closedloopcontrolsystem}
	\end{equation}
	Solutions to system $\Sigma_u$  are given by the trajectory $t\in[t_0,T]\mapsto x(t)=\psi_{t_0}^u(t,x_0)$   from the initial condition $x_0\in\mathcal{X}$, for a fixed initial $u_0\in\mathcal{U}$, at time $t_0$, with $\psi_{t_0}^{u_0}(t_0,x_0)=x_0$.  Consider a  simply connected  neighborhood $\mathcal{C}$ of $\mathcal{X}$  such that $\psi_{t_0}^{u}(t,x_0)$ is forward complete for every $x_0\in\mathcal{C}$, i.e.,  $\psi_{t_0}^u(t,x_0)\in\mathcal{C}$ for each $t_0$, each $u_0$ and each $t \geq t_0$. Solutions to $\Sigma$ are defined in a similar manner and are denoted by  $x(t)=\psi_{t_0}(t,x_0)$. By  connectedness of $\mathcal{C}$, any two points in $\mathcal{C}$ can be connected by a regular smooth curve  $\gamma:I\rightarrow \mathcal{C}$, with $I:=[0,1]$. A function $\alpha:\mathds{R}_{\geq 0}\rightarrow\mathds{R}_{\geq 0}$ is said to be of class $\mathcal{K}$ if it is strictly increasing and	$\alpha(0) = 0$ \cite{khalil1996noninear}. 	 When it is clear from the context, some function arguments  will be left out in the rest of this paper.
	
	\begin{defin}[Incremental stability \cite{forni}]\label{definition:incrementalstability}
		Let $\mathcal{C}\subseteq\mathcal{X}$ be a forward invariant set,  $d:\mathcal{X}\times\mathcal{X}\rightarrow \mathds{R}_{\geq 0}$ be a continuous metric and consider system $\Sigma$   given by \eqref{eq:closedloopcontrolsystem}. Then, system $\Sigma$ is said to be
		\begin{itemize}
			\item  \emph{Incrementally stable} ($\Delta$-S) on $\mathcal{C}$  (with respect to $d$) if there exist a $\mathcal{K}$ function   $\alpha$ such that for each $x_1, x_2\in\mathcal{C}$, for each $t_0\in\mathds{R}_{\geq 0}$ and for  all $t\geq t_0$,
			\begin{equation}
			d(\psi_{t_0}(t,x_1),\psi_{t_0}(t,x_2)) \leq \alpha (d(x_1,x_2)).
			\end{equation}
			\item  \emph{Incrementally asymptotically  stable} ($\Delta$-AS) on $\mathcal{C}$ if it is $\Delta$-S and for all $x_1,x_2\in\mathcal{C}$, and for each $t_0\in\mathds{R}_{\geq 0}$,
			\begin{equation}
			\lim\limits_{t\rightarrow\infty}d(\psi_{t_0}(t,x_1),\psi_{t_0}(t,x_2)) =0.
			\end{equation}
			\item  \emph{Incrementally exponentially stable} ($\Delta$-ES) on $\mathcal{C}$ if there exist a distance $d$, $k \geq 1$, and $\beta>0$ such that for each $x_1, x_2\in\mathcal{C}$, fir each $t_0\in\mathds{R}_{\geq 0}$ and for all $t\geq t_0$,
			\begin{equation}
			d(\psi_{t_0}(t,x_1),\psi_{t_0}(t,x_2)) \leq k e^{-\beta(t-t_0)}d(x_1,x_2).
			\end{equation}
		\end{itemize}
	\end{defin}
	Above definitions are the   incremental versions of the classical notions of stability, asymptotic stability and exponential stability \cite{khalil1996noninear}. If   $\mathcal{C}=\mathcal{X}$,  then  we say global $\Delta$-S, $\Delta$-AS and $\Delta$-ES, respectively. All properties are assumed to be uniform in $t_0$.
	
	\subsubsection{Differential Lyapunov theory and contraction analysis}			
	\begin{defin}
		The \emph{prolonged\cite{crouch} control system}  $\Sigma_u^{\delta}$  associated to the control system $\Sigma_u$ in  \eqref{eq:controlsystem} is given by
		\begin{equation}
		\begin{split}
		\Sigma_u^{\delta}:\left\{ 
		\begin{array}{lcc}
		\dot{{x}}={f}({x},t)+\sum_{i=1}^{n}{g}_i({x},t){u}_i,\\
		y=h(x,t),\\
		\delta\dot{x}=\frac{\partial f}{\partial x}(x,t)\delta x+\sum_{i=1}^{n}u_i\frac{\partial g_i}{\partial x}(x,t)\delta x  +\sum_{i=1}^{n}g_i(x,t)\delta u_i,\\
		\delta y=\frac{\partial h}{\partial x}(x,t)\delta x.\\
		\end{array}
		\right.
		\end{split}
		\label{eq:prolongedcontrolsystem}
		\end{equation}
		with    $({u},\delta{u})\in T\mathcal{U}$,   $({x},\delta{x})\in T\mathcal{X}$, and  $({y},\delta{y})\in T\mathcal{Y}$. The  \emph{prolonged system} $\Sigma^{\delta}$ of $\Sigma$ in   \eqref{eq:closedloopcontrolsystem} is  similarly defined as
		\begin{equation}
		\begin{split}
		\Sigma^{\delta}:\left\{ \begin{array}{lcc}
		\dot{{x}}={F}({x},t), \\
		{y}={h}(x,t),\\
		\delta\dot{{x}}=\frac{\partial {F}}{\partial{x}}({x},t)\delta{x},\\
		\delta{y}=\frac{\partial {h}}{\partial x}(x,t)\delta x.\\		
		\end{array}
		\right.
		\end{split}
		\label{eq:prolonged-closedloop-system}
		\end{equation}		
	\end{defin}

	\begin{defin}\label{defin:finslerlyapunov}
		A function  $V:T\mathcal{X}\times \mathds{R}_{\geq 0}\rightarrow \mathds{R}_{> 0}$ is a candidate \emph{differential or Finsler-Lyapunov function} if it satisfies 
		\begin{equation}
		c_1 \mathcal{F}({x},\delta{x},t)^{p} \leq V({x},\delta{x},t) \leq c_2\mathcal{F}({x},\delta{x},t)^{p},
		\label{eq:finslerlyapunov}
		\end{equation}
		for some   $c_1,c_2\in\mathds{R}_{>0}$, and with $p$   a positive integer where $\mathcal{F}({x},\delta x,t)$ is a Finsler structure\cite{forni},  uniformly in $x$ and $t$.
	\end{defin}
	The relation between a candidate differential Lyapunov function and the Finsler structure in \eqref{eq:finslerlyapunov} is a key property for incremental stability analysis,  since it implies the existence of a  well-defined distance on $\mathcal{X}$ via integration as defined below.	
	\begin{defin}\label{defin:finslerdistance}
		Consider a candidate differential Lyapunov function on   $\mathcal{X}$ and the associated Finsler structure $\mathcal{F}$. For any subset $\mathcal{C}\subseteq\mathcal{X}$ and any $x_1,x_2\in\mathcal{C}$, let $\Gamma({x}_1,{x}_{2})$ be the collection of piecewise $C^1$ curves ${\gamma}:I\rightarrow \mathcal{X}$ connecting $x_1$ and $x_2$ with ${\gamma}(0)={x}_{1}$ and ${\gamma}(1)={x}_{2}$. \emph{The Finsler distance} $d:\mathcal{X}\times\mathcal{X}\rightarrow\mathds{R}_{\geq 0}$ induced by the structure $\mathcal{F}$ is defined by
		\begin{equation}
		d({x}_1,{x}_2):=\inf_{\Gamma({x}_1,{x}_2)}\int_{\gamma}{\mathcal{F}\left({\gamma}(s),\frac{\partial {\gamma}}{\partial s}(s),t\right)}ds.
		\label{defin:finslerdistance1}
		\end{equation}
	\end{defin}	
	The following result gives a sufficient condition for incremental stability in terms of  differential Lyapunov functions.
	\begin{theorem}[Direct differential Lyapunov method \cite{forni}]\label{theo:lyapunovcontraction}
		Consider the prolonged system $\Sigma^{\delta}$ in \eqref{eq:prolonged-closedloop-system}, a connected and forward invariant set $\mathcal{C}\subseteq\mathcal{X}$, and a function $\alpha:\mathds{R}_{\geq 0}\rightarrow \mathds{R}_{\geq 0}$. Let $V$ be a candidate differential Lyapunov function satisfying  
		\begin{equation}
		\dot{V}({x},\delta {x},t)	\leq -\alpha(V({x},\delta {x},t)) 
		\label{eq:finslerlyapunovinequality}
		\end{equation}
		for each $({x},\delta{x},t)\in T\mathcal{X}\times\mathds{R}_{\geq 0}$ uniformly in $t$. Then, system $\Sigma$ in \eqref{eq:closedloopcontrolsystem} is 
		\begin{itemize}
			\item incrementally stable on $\mathcal{C}$ if $\alpha(s)=0$ for each $s\geq 0$;
			\item  Incrementally asymptotically stable  on $\mathcal{C}$ if $\alpha$ is a $\mathcal{K}$ function;
			\item  incrementally exponentially stable on $\mathcal{C}$ if $\alpha(s)=\beta s, \forall s>0$.
		\end{itemize}			 
	\end{theorem}
	
	\begin{defin}\label{def:contraction}
		We say that $\Sigma$  \emph{contracts\cite{forni} (respectively does not expand\cite{sanfeliceconvergence2} )} $V$ in $\mathcal{C}$ if \eqref{eq:finslerlyapunovinequality} is satisfied for a function $\alpha$ of class $\mathcal{K}$ (resp.  $\alpha(s)=0$ for all $s\geq 0$). The set  $\mathcal{C}$ is  \emph{the contraction region} (resp.  \emph{nonexpanding region}).
	\end{defin}
	
	%

	\begin{remark}[Riemannian contraction metrics]\label{remark:generalizedcontractionnalaysis}
		The so-called \emph{generalized contraction analysis} in Lohmiller\cite{slotinecontraction} with Riemannian metrics can be seen as a particular case of  Theorem \ref{theo:lyapunovcontraction} as follows: Take as candidate differential Lyapunov function to 
		\begin{equation}
		V({x},\delta{x},t)=\frac{1}{2}\delta{x}^{\top}{\Pi}({x},t)\delta{x},
		\label{eq:SlotinedifferentialLyapu}
		\end{equation}
		where  $\mathcal{F}(x,\delta x,t)=\sqrt{V(x,\delta x,t)}$, ${\Pi({x},t)}=\Theta^{\top}(x,t)\Theta(x,t)$   and $\Theta:\mathcal{X}\times\mathds{R}_{\geq 0}\rightarrow \mathds{R}^{N\times N}$ is smooth and positive for all $t$. If 
		\begin{equation}
		\dot{\Pi}(x,t)+\frac{\partial {F}^{\top}}{\partial{x}}{\Pi}({x},t)+{\Pi}({x},t)\frac{\partial {F}}{\partial {x}}\leq -2\beta{\Pi}(x,t).
		\label{eq:slotinecontractionconditions}
		\end{equation}
		holds for all $x\in\mathcal{X}$, uniformly in $t$, then, $\Sigma$ contracts \eqref{eq:SlotinedifferentialLyapu}. 
		Condition \eqref{eq:slotinecontractionconditions} is equivalent to verify that the \emph{generalized Jacobian}\cite{slotinecontraction} 
		\begin{equation}
		\overline{J}(x,t)=\left[\dot{\Theta}(x,t)F(x,t)+\Theta(x,t)\frac{\partial {F}}{\partial{x}}\right]\Theta^{-1}(x,t).
		\label{eq:generalizedJacobian}
		\end{equation}
		satisfies\cite{forni,coogan2017contractive} $\mu(\overline{J}(x,t))\leq -2\beta$ uniformly in $t$, where $\mu(\cdot)$ is a  matrix measure\footnote{Given a vector norm $|\cdot|$ on a linear space, with its induced matrix norm $\|A\|$, the associated matrix measure $\mu$ is defined\cite{sontag2010contractive} as the directional derivative of the matrix norm in the direction of $A$ and evaluated at the identity matrix, that is: \textcolor{blue}{$\mu(A):=\lim\limits_{h\rightarrow 0}\frac{1}{h}\left(\|I_n+hA\|-1\right)$, where $I_n$ is the $n\times n$ identity matrix}.} as shown by Russo \cite{russo2010global}, Forni \cite{forni} and Coogan \cite{coogan2017contractive}. 
	\end{remark}	
	
	

	\subsubsection{Differential passivity}
	
	
	\begin{defin}[van der Schaft\cite{arjan2013differentialpassivity}, Forni\cite{forni2013differentially}]\label{def:differential-passivity}
		Consider a nonlinear control system $\Sigma_u$ in \eqref{eq:controlsystem} together with its prolonged system $\Sigma_u^{\delta}$ given by \eqref{eq:prolongedcontrolsystem}. Then, $\Sigma_u$  is called \emph{differentially passive} if the prolonged system $\Sigma_u^{\delta}$  is dissipative with respect to the supply rate $\delta y^{\top}\delta u$,  i.e., if there exist a \emph{differential storage function} function ${W}:T\mathcal{X}\times\mathds{R}_{\geq 0}\rightarrow\mathds{R}_{\geq 0} $ satisfying 
		\begin{equation}
		\frac{d W}{dt}(x,\delta x,t)\leq \delta y^{\top}\delta u,
		\label{eq:differentialpassivityinequality}
		\end{equation}
		for all $x,\delta x,u,\delta u$ uniformly in $t$. Furthermore,  system \eqref{eq:controlsystem} is called \emph{differentially lossless} if \eqref{eq:differentialpassivityinequality} holds with equality.
	\end{defin}	
	If additionally, the differential storage function is required to be a differential Lyapunov function, then differential passivity implies contraction when the variational input is $\delta u=0$. For further details we refer to the works of van der Schaft\cite{arjan2013differentialpassivity} and Forni \cite{forni2013differential}. 
	
	The following lemma characterizes the structure of a class of control systems   which are  differentially passive.
	\begin{lemma}[Reyes-B\'aez\cite{RodoAutomatica2017}]\label{lemma:dPBC}
		Consider the control system $\Sigma_u$ in \eqref{eq:controlsystem} together with its prolonged system $\Sigma_u^{\delta}$ in \eqref{eq:prolongedcontrolsystem}. Suppose there exists a  transformation $\delta \tilde{x}=  \Theta(x,t)\delta x$ such that the variational dynamics  in \eqref{eq:prolongedcontrolsystem} given by
		\begin{equation}
		\begin{split}
		\delta\Sigma_u:\left\{ 
		\begin{array}{lcc}
		\delta\dot{x}=\frac{\partial f}{\partial x}(x,t)\delta x+\sum_{i=1}^{n}u_i\frac{\partial g_i}{\partial x}(x,t)\delta x  +\sum_{i=1}^{n}g_i(x,t)\delta u_i,\\
		\delta y=\frac{\partial h}{\partial x}(x,t)\delta x,\\
		\end{array}
		\right.
		\end{split}
		\label{eq:prolongedcontrolsystem-variational}
		\end{equation}		
		takes the form
		\begin{equation}
		\begin{split}
		\delta\tilde{\Sigma}_u:\left\{ 
		\begin{array}{lcc}
		\delta \dot{\tilde{x}}=\left[\Xi(\tilde{x},t)-\Upsilon(\tilde{x},t)\right]\Pi(\tilde{x},t)\delta\tilde{x}+\Psi(\tilde{x},t)\delta u,\\
		\delta\tilde{y}=\Psi^{\top}(\tilde{x},t)\Pi(\tilde{x},t)\delta \tilde{x},
		\end{array}
		\right.
		\end{split}
		\label{eq:feedbackinterconnectionsystemsvariational}	
		\end{equation}
		where   $\Pi(\tilde{x},t)>0_N$ is a Riemannian metric tensor,   $\Xi(\tilde{x},t)=-\Xi^{\top}(\tilde{x},t)$, $\Upsilon(\tilde{x},t)$ are rectangular matrices. If condition 
		\begin{equation}
		\delta\tilde{x}^{\top}\left[\dot{\Pi}(\tilde{x},t)-\Pi(\tilde{x},t) (\Upsilon(\tilde{x},t)+\Upsilon^{\top}(\tilde{x},t))\Pi(\tilde{x},t)\right]\delta\tilde{x}\leq-\alpha(W(\tilde{x},\delta \tilde{x},t)),
		\label{eq:dPBCinequality}
		\end{equation}
		holds for all $(\tilde{x},\delta \tilde{x})\in T\mathcal{X}$ uniformly in $t$, with  $\alpha$   of class $\mathcal{K}$. Then, $\Sigma_u$ is differentially passive from   $\delta u$ to $\delta\tilde{y}$ with respect to the differential storage function  given by
		\begin{equation}
		W(\tilde{x},\delta \tilde{x},t)=\frac{1}{2}\delta \tilde{x}^{\top}{\Pi}(\tilde{x},t)\delta \tilde{x}.
		\label{eq:differentialStorage-Lemma}
		\end{equation}			
	\end{lemma}
	The passivity theorem of negative feedback interconnection of two passive systems resulting in a passive closed-loop system can be extended to differential passivity as follows. Consider two differentially passive nonlinear systems $\Sigma_{u_i}$, with states $x_i\in\mathcal{X}_i$, inputs $u_i\in\mathcal{Y}_i$, outputs $u_i\in\mathcal{U}$ and differential storage functions $W_i$, for $i\in\{1,2\}$. The standard feedback interconnection is
	\begin{equation}
	u_1=-y_2+e_1,\quad u_2=y_1+e_2,
	\label{eq:feedback-interconnection}
	\end{equation}
	where $e_1,e_2$ denote external outputs. The equations \eqref{eq:feedback-interconnection} imply that the variational quantities $\delta u_1,\delta u_2, \delta y_1, \delta y_2, \delta e_1, \delta e_2$ satisfy
	\begin{equation}
	\delta u_1=-\delta y_2+\delta e_1,\quad \delta u_2=\delta y_1+\delta e_2.
	\label{eq:feedback-interconnection-variational}
	\end{equation}
	The variational feedback interconnection \eqref{eq:feedback-interconnection-variational} implies that the  equality  $\delta u_1^{\top} \delta y_1+\delta u_2^{\top} \delta y_2=\delta e_1^{\top} \delta y_1+\delta e_2^{\top} \delta y_2$ holds.
	Thus, the closed-loop system arising from the feedback interconnection in \eqref{eq:feedback-interconnection-variational} of $\Sigma_{u_1}$ and $\Sigma_{u_2}$ is a differentially passive system with supply rate $\delta e_1^{\top} \delta y_1+\delta e_2^{\top} \delta y_2$ and storage function $W=W_1+W_2$, as it is shown by van der Schaft \cite{arjan2013differentialpassivity}.
	
	\subsubsection{Contraction and differential passivity of virtual systems}	
	\begin{defin}[Reyes-Baez\cite{RodoAutomatica2017}, Wang\cite{slotinecontraction}]\label{defin:virtualsystem1}
		Consider systems $\Sigma_u$ and $\Sigma$, given by  \eqref{eq:controlsystem} and \eqref{eq:closedloopcontrolsystem}, respectively. Suppose that  $\mathcal{C}_v\subseteq \mathcal{X}$ and $\mathcal{C}_x\subseteq \mathcal{X}$ are connected and forward invariant. 
		A \emph{virtual control system}  associated to $\Sigma_u$ is defined as 
		\begin{equation}
		\begin{split}
		{\Sigma}_{u}^v:\left\{ \begin{array}{lcc}
		\dot{x}_v=\Gamma_v(x_v,x,u_v,t),\\		{y}_v={h}_v(x_v,x,t), \quad\quad  \forall t\geq t_0,		
		\end{array}
		\right.			
		\end{split}
		\label{eq:virtualsystemTheorem1}	
		\end{equation}
		with state  $x_v\in\mathcal{X}$ and parametrized by   $x\in\mathcal{X}$, where      $\Gamma_v:\mathcal{C}_v\times\mathcal{C}_x\times\mathcal{U}\times\mathds{R}_{\geq 0}\rightarrow T \mathcal{X}$ and $h_v:\mathcal{C}_v\times\mathcal{C}_x\times\mathds{R}_{\geq 0}\rightarrow  \mathcal{Y}$ are such that 
		\begin{equation}
		\begin{split}
		\Gamma(x,x,u,t)=f(x,t)+\sum_{i=1}^{n}{g}_i(x,t){u}_i,\quad
		h_v(x,x,t)=h(x,t);\quad\quad\quad \forall u, \forall t\geq t_0.		
		\end{split}
		\label{eq:virtualcontrolsystemTheorem1}		
		\end{equation}		
		Similarly, a  \emph{virtual system} associated to $\Sigma$ is defined as 
		\begin{equation}
		{\Sigma}^{v}:\left\{ \begin{array}{lcc}
		\dot{{x}}_v= {\Phi}_v(x_v,{x},t), \\
		{y}_v={h}_v(x_v,x,t),		
		\end{array}
		\right.		
		\label{eq:virtualsystemTheorem}
		\end{equation}
		with state $x_v\in\mathcal{C}_v$ and  parametrized by $x\in\mathcal{C}_x$, where   $\Phi_v:\mathcal{C}_v\times\mathcal{C}_x\times\mathds{R}_{\geq 0}\rightarrow T\mathcal{X}$ and $h_v:\mathcal{C}_v\times\mathcal{C}_x\times\mathds{R}_{\geq 0}\rightarrow \mathcal{Y}$  satisfying 
		\begin{equation}
		{\Phi}_v({x},{x},t)=F(x,t)\quad \text{and}\quad h_v(x,x,t)=h(x,t), \quad\quad \text{for all} t>t_0.
		\label{eq:virtualsystemTheorem-condition}		
		\end{equation}
	\end{defin}	
	It follows that any solution $x(t)=\psi_{t_0}(t,x_o)$ of the \emph{actual control system}  $\Sigma_u$  in \eqref{eq:controlsystem},  starting at $x_0\in\mathcal{C}_x$  for a certain input ${u}$, generates the solution $x_v(t)=\psi_{t_0}(t,x_0)$ to the virtual system  $\Sigma_{u}^v$ in \eqref{eq:virtualsystemTheorem1}, starting at $x_{v0}=x_0\in\mathcal{C}_v$ with $u_v=u$, for all $t>t_0$. In a similar manner for the closed actual system $\Sigma$ in \eqref{eq:closedloopcontrolsystem}, any solution $x(t)=\psi_{t_0}(t,x_o)$ starting at $x_0\in\mathcal{C}_x$, generates the solution $x_v(t)=\psi_{t_0}(t,x_o)$ to the closed virtual system $\Sigma^v$ in \eqref{eq:virtualsystemTheorem}, starting at $x_{v0}=x_0\in\mathcal{C}_v$,  for all $t>t_0$. However, \emph{not} every virtual system's solution $x_v(t)$  corresponds to an  actual system's solution. Thus, \emph{for \emph{any} trajectory $x(t)$, we may consider \eqref{eq:virtualsystemTheorem1} (respectively \eqref{eq:virtualsystemTheorem}) as a time-varying system with state $x_v$.}

	\begin{theorem}[Virtual contraction \cite{wang,forni}]\label{theo:partialcontraction}
		Consider  systems $\Sigma$ and $\Sigma^v$ given by \eqref{eq:closedloopcontrolsystem} and \eqref{eq:virtualsystemTheorem}, respectively. Let  $\mathcal{C}_v\subseteq \mathcal{X}$ and $\mathcal{C}_x\subseteq \mathcal{X}$ be two connected and forward invariant sets. Suppose that $\Sigma^v$  is uniformly contracting with respect to $x_v$. Then, for any initial conditions $x_0\in\mathcal{C}_x$ and $x_{v0}\in\mathcal{C}_v$, each solution to $\Sigma^v$ converges asymptotically  to the solution  of $\Sigma$.
	\end{theorem}	
	If the conditions of Theorem \ref{theo:partialcontraction} hold, then system $\Sigma$ is said to be \emph{virtually contracting}.  		
	If   the virtual   system $\Sigma_{u}^v$ is differentially passive,  then   the    system $\Sigma_u$ is said to be \emph{virtually differentially passive}. In this case, the steady-state solution is driven by the input and is denoted by $\overline{x}_v^{u_v}(t)=x^u(t)$.   This last property can be used for v-CBC, as will be shown later.

	\subsubsection{Virtual contraction based control (v-CBC)}\label{subsection:v-dPBC}				
	From a control design point of view, the usual task is to render a specific solution of the system   exponentially/asymptotically stable,  rather than the stronger contractive behavior of all system's solutions. In this regard, as an alternative to the existing control techniques in the literature, we propose a design method based on the concept of virtual contraction to solve the set-point regulation or trajectory tracking problems. Thus, the control objective  is to design a  scheme such that a   well-defined Finsler distance  between  the solution starting at $t_0$ and desired solution shrinks by means of virtual system's contracting behavior.\\
	
	The proposed  design methodology is divided in three main steps:
	\begin{enumerate}
		\item Propose a  virtual   system  \eqref{eq:virtualsystemTheorem1} for system  \eqref{eq:controlsystem}.
		\item Design a  state feedback $u_v=\zeta(x_v,x,t)$ for the virtual system \eqref{eq:virtualsystemTheorem1}, such that the closed-loop   system   is contractive  and tracks a predefined reference solution.
		\item Define the controller for the actual  system \eqref{eq:controlsystem} as $u=\zeta(x,x,t)$.
	\end{enumerate}	
	If we are able to design a controller with the above steps, then, according to Theorem \ref{theo:partialcontraction},  all the solutions of the closed-loop virtual system will converge to the closed-loop original system solution starting at $x_0$, that is, $\overline{x}(t)=x_d(t)\rightarrow x(t)$  as $t\rightarrow \infty$.\\
	
	\subsection{A class of virtual control systems for mechanical systems in the port-Hamiltonian framework}
	In this subsection, the previous notions on contraction and differential passivity are applied   to mechanical systems described in the port-Hamiltonian  framework\cite{vanderschaft1995}.
	\subsubsection{Port-Hamiltonian formulation of mechanical systems}	
	\begin{definition}
		A {port-Hamiltonian system} with  $N$ dimensional state space manifold $\mathcal{X}$, input and output spaces $\mathcal{U}=\mathcal{Y}\subset\mathds{R}^{m}$, and Hamiltonian function $H:\mathcal{X}\rightarrow\mathds{R}$, is given by
		\begin{equation}
		\begin{split}
		\dot{x}&=\left[J(x)-R(x)\right]\frac{\partial H}{\partial x}(x)+g(x)u\\			
		y&= g^{\top}(x)\frac{\partial H}{\partial x}(x),
		\end{split}
		\label{eq:IOpHsystem}			
		\end{equation}
		where $g(x)$ is a $N\times m$  matrix,    $J(x)=-J^{\top}(x)$ is the $N\times N$ interconnection matrix and $R(x)=R^{\top}(x)$ is the $N\times N$ positive semi-definite dissipation matrix. 
	\end{definition}
	
	In the specific case of a mechanical system with generalized coordinates $q$ on the configuration space $\mathcal{Q}$ of dimension $n$ and  velocity $\dot{{q}}\in T_{q}\mathcal{Q}$, the Hamiltonian function is given by the total energy
	\begin{equation}
	H({q,p})=\frac{1}{2}{p}^{\top}{M}^{-1}({q}){p}+P({q}),
	\label{eq:phmechanicalHamiltonian}
	\end{equation}
	where ${x}=({q},{p})\in T^*\mathcal{Q}$ is the   state, $P({q})$ is the potential energy,  $p:=M(q)\dot{q}$ is the   momentum and the inertia matrix $M(q)$ is symmetric and positive definitive. Then,  the pH system \eqref{eq:IOpHsystem} takes the form
	\begin{equation}
	\begin{split}
	\begin{bmatrix}
	\dot{q}\\
	\dot{p}
	\end{bmatrix}&=\begin{bmatrix}
	0_n & I_v\\
	-I_n & -D(q,p)
	\end{bmatrix}\begin{bmatrix}
	\frac{\partial H}{\partial q}(q,p)\\
	\frac{\partial H}{\partial p}(q,p)
	\end{bmatrix}+\begin{bmatrix}
	0_n \\
	B(q)
	\end{bmatrix}u,\\
	y&=B^{\top}(q)\frac{\partial H}{\partial p}(q,p),
	\end{split}
	\label{eq:Hamiltonian2}
	\end{equation}
	with matrices
	\begin{equation}
	J(x)=\begin{bmatrix}
	{0}_n & {I}_n\\
	-{I}_n& {0}_n
	\end{bmatrix};\quad\quad  R(x)=\begin{bmatrix}
	{0}_n& {0}_n\\
	{0}_n& {D}({{q}},p)
	\end{bmatrix};\quad\quad g(x)=\begin{bmatrix}
	{0}_n\\
	B(q)
	\end{bmatrix},
	\label{eq:phmechanical}		
	\end{equation}		
	where  ${D}({q},p)={D}^{\top}({q},p) \geq {0}_n$ is the damping matrix and $I_n$ and $0_n$ are the $n\times n$ identity, respectively, zero matrices. The input force matrix $B(q)$ has rank $m\leq n$; if $m< n$ we say that the mechanical system is underactuated, otherwise it is fully-actuated. System \eqref{eq:Hamiltonian2} defines the passive  map $u\mapsto y$ with respect to the Hamiltonian   \eqref{eq:phmechanicalHamiltonian} as  storage function.
	
	Using the structure of the internal workless forces, system \eqref{eq:Hamiltonian2} can be equivalently rewritten as, see Reyes-B\'aez\cite{reyes2017virtual,ReyesBaez-PhDthesis}.
	\begin{equation}
	\begin{split}
	\begin{bmatrix}
	\dot{{q}}\\
	\dot{{p}}
	\end{bmatrix}&=\begin{bmatrix}
	{0}_n & {I}_n \\
	-{I}_n & -({E}(q,p)+D(q,p))
	\end{bmatrix}\begin{bmatrix}
	\frac{\partial P}{\partial {q}}(q)  \\
	\frac{\partial H}{\partial {p}}(q,p)
	\end{bmatrix}+\begin{bmatrix}
	{0}_n \\
	B(q)
	\end{bmatrix}{u},\\
	y_E&=\begin{bmatrix}
	{0}_n & {B}^{\top}(q)
	\end{bmatrix}\begin{bmatrix}
	\frac{\partial P}{\partial {q}}(q)  \\
	\frac{\partial H}{\partial {p}}(q,p)
	\end{bmatrix},
	\end{split}
	\label{eq:phmechanical-alternative}
	\end{equation}	
	where  ${E}(q,p):=S_H(q,p)-\frac{1}{2}\dot{{M}}({q})$, and $S_H(q,p)={S}_L(q,\dot{q})|_{\dot{q}=M^{-1}(q)p}$ is a skew-symmetric matrix whose $(k,j)$-th element   is\footnote{The structure of matrix $S_L(q,\dot{q})$ is a consequence of the fact that Hamilton's principle is satisfied. This was first reported  by Arimoto and Miyazaki \cite{Arimoto84}.}
	\begin{equation}
	\begin{split}
	S_{Lkj}(q,\dot{q})
	&= \frac{1}{2} \sum_{i=1}^{n}\left\{\frac{\partial M_{ki}}{\partial q_j}(q)  -\frac{\partial  M_{ij}}{\partial q_k}(q) \right\}\dot{q}_i.
	\label{eq:SlLagrangian}
	\end{split}
	\end{equation} 	
	From the energy balance  along the trajectories of   \eqref{eq:phmechanical-alternative}, it is easy to see that forces $E(q,p)M^{-1}(q)p$ are \emph{workless}, i.e., their power is zero. Thus, system \eqref{eq:phmechanical-alternative} preserves the passivity property of the map $u\mapsto y=y_E$, as well with  \eqref{eq:phmechanicalHamiltonian} as storage function.

	\subsubsection{A class of virtual control systems for mechanical pH systems}
	Let $x=[q^{\top},p^{\top}]^{\top}\in T^*\mathcal{Q}$ be the state of system \eqref{eq:Hamiltonian2}.  Following Definition \ref{defin:virtualsystem1} and considering the port-Hamiltonian formulation  \eqref{eq:phmechanical-alternative} of \eqref{eq:Hamiltonian2}, we construct the virtual mechanical control system  associated to \eqref{eq:Hamiltonian2}  as the time-varying system given by\cite{reyes2017virtual}
	\begin{equation}
	\begin{split}
	\dot{x}_v&=\begin{bmatrix}
	{0}_n & {I}_n\\
	-{I}_n& -({E}(x)+D(x))
	\end{bmatrix}\begin{bmatrix}
	\frac{\partial H_v}{\partial {q}_v}({x_v,x})\hfill \\
	\frac{\partial H_v}{\partial {p}_v}(x_v,x)
	\end{bmatrix}+\begin{bmatrix}
	{0}_n\\
	{B}(q)
	\end{bmatrix}{u}_v\\
	y_v&=\begin{bmatrix}
	{0}_n& {B}^{\top}({q})
	\end{bmatrix}\begin{bmatrix}
	\frac{\partial H_v}{\partial {q}_v}({x_v,x})\hfill \\
	\frac{\partial H_v}{\partial {p}_v}(x_v,x)
	\end{bmatrix},
	\end{split}
	\label{eq:phmechanicalvirtual}
	\end{equation}
	with state $x_v=(q_v, p_v)\in \mathcal{X}$,  parametrized by the state trajectory   $x(t)$  of \eqref{eq:phmechanical-alternative},
	and with Hamiltonian-like function
	\begin{equation}
	H_v(x_v,x)=\frac{1}{2}p_v^{\top}M^{-1}(q)p_v+P_v(q_v).
	\label{eq:phmechanicalHamiltonianvirtual}
	\end{equation}	
	where $P_v(q_v):=P(q_v)$.  Remarkably, the virtual control system \eqref{eq:phmechanicalvirtual} is also  passive with   input-output pair $(u_v,y_v)$ and  $x$-parametrized  storage function  \eqref{eq:phmechanicalHamiltonianvirtual}, for every state trajectory  $x(t)$ of \eqref{eq:phmechanical-alternative}. Furthermore, system \eqref{eq:phmechanicalvirtual} can be rewritten as 
	
	\begin{equation}
	\begin{split}
	\dot{x}_v&=\left[J_v(x)-R_v(x)\right]\frac{\partial H_v}{\partial x_v}(x_v,x)+g(x){u}\\
	y_v&=g^{\top}(x)\frac{\partial H_v}{\partial x_v}(x_v,x),
	\end{split}
	\label{eq:pHlike-virtual}
	\end{equation}
	with $g(x)$   as in \eqref{eq:phmechanical} and matrices
	\begin{equation}
	J_v(x)=\begin{bmatrix}
	0_n & I_n \\
	-I_n  & -S_H(x)
	\end{bmatrix},\quad\quad \ R_v(x):=\begin{bmatrix}
	0_n  & 0_n \\
	0_n  & (D(x)-\frac{1}{2}\dot{M}(x))
	\end{bmatrix}, 
	\label{eq:pH-likesystem}
	\end{equation}	
	where $J_v(x)=-J_v^{\top}(x)$  and $R_v(x)=R^{\top}_v(x)$. The skew-symmetric matrix $J_v(x)$ defines an \emph{almost-Poisson tensor\cite{ReyesBaez-PhDthesis}} implying that   energy conservation  is satisfied.  However,  system \eqref{eq:pHlike-virtual} is not a pH system since  $R_v(x)\geq 0$ does not necessarily hold. Thus, we refer to system \eqref{eq:pHlike-virtual}  as a \emph{mechanical pH-like system}.	
	The variational virtual dynamics of system \eqref{eq:pHlike-virtual} is
	\begin{equation}
	\begin{split}
	\delta \dot{x}_v&=\left[J_v(x)-R_v(x)\right]\frac{\partial^2 H_v}{\partial x_v^2}(x_v,x)\delta x_v+g(x)\delta u		\\
	\delta y_v&=g^{\top}(x)\frac{\partial^2 H_v}{\partial x_v^2}(x_v,x)\delta x_v.
	\end{split}
	\label{eq:variationalvirtualpHsystem}
	\end{equation}	
	Notice that   \eqref{eq:variationalvirtualpHsystem} is of the form   \eqref{eq:feedbackinterconnectionsystemsvariational} with $\Xi(x_v,t)=J_v(x)$, $\Upsilon(x_v,t)=R_v(x)$ and $\Pi(x_v,t)=\frac{\partial^2 H_v}{\partial x_v^2}(x_v,x)$.   Moreover, if   hypotheses in Lemma \ref{lemma:dPBC} are satisfied, then   system \eqref{eq:phmechanicalvirtual} is \emph{differentially passive} with supply rate $\delta y^{\top}\delta u$.	
	\section{Problem Statement}\label{sec:ProbStat}
	
	\subsection{Flexible-joints robots as port-Hamiltonian systems}
\begin{figure}[h!]
		\centering		
		\includegraphics[width=7cm]{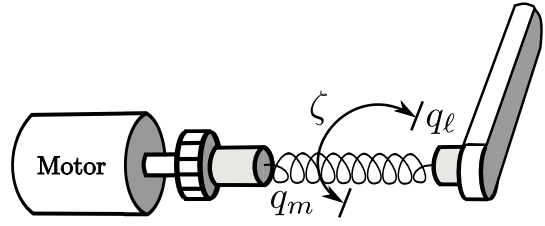}
		\caption{Flexible joint mechanical structure: motor's shaft position $q_m$, spring's deflection $\zeta$ and link's position $q_{\ell}$.}
		\label{fig:flexiblejoint}		
	\end{figure}

	FJRs are a class of robot manipulators in which each  joint  is given by a link interconnected to a motor through a spring; see Figure \ref{fig:flexiblejoint}. Two generalized coordinates are needed to describe the configuration of a single flexible-joint, these are given by the link $q_{\ell}$ and motor $q_m$  positions as shown in Figure \ref{fig:flexiblejoint}. 

	Thus, FJRs are a class of \emph{underactuated} mechanical systems of  $n=\text{dim}\mathcal{Q}$ degrees of freedom (dof). The dof corresponding to the $n_m$-motors position are actuated, while the dof corresponding to the $n_{\ell}=n_m$   links position are underactuated, with $n=n_{m}+n_{\ell}$. We consider the following standard modeling assumptions in Spong \cite{spongflexible} and Jard\'on-Kojakhmetov\cite{hilde}:
	\begin{itemize}
		\item The deflection/elongation  $\zeta$ of each spring is small enough so that it is represented by a linear model.
		\item The $i$-th motor driving the $i$-link is mounted at the $(i-1)$-link.
		\item Each motor's center of mass is located along the rotation axes.
	\end{itemize}
	The FJR's generalized position $q\in\mathcal{Q}$  is split as $q=[q_{\ell}^{\top},q_m^{\top}]^{\top}\in\mathcal{Q}=\mathcal{Q}_{n_{\ell}}\times \mathcal{Q}_{n_m}$,  the inertia and damping matrices are assumed to be block partitioned as  follows
	\begin{equation}
	M(q)=\begin{bmatrix}
	M_{\ell}(q_{\ell}) & 0_{n_{\ell}}\\
	0_{n_m} & M_m(q_m)
	\end{bmatrix}; \quad\quad D(x)=\begin{bmatrix}
	D_{\ell}(q_{\ell},p_{\ell}) & 0_{n_{\ell}}\\
	0_{n_m} & D_m(q_m,p_m)
	\label{label:kinetic-flexible}
	\end{bmatrix},
	\end{equation}
	where $M_{\ell}(q_{\ell})$ and $M_m(q_m)$ are the link and motors inertia matrices, and  $D_{\ell}(q_{\ell},p_{\ell})$ and $D_m(q_m,p_m)$ are the link and motor damping matrices. The total potential energy is given by
	\begin{equation}
	P(q)=P_{\ell g}(q_{\ell })+P_{m g}(q_m)+ P_{\zeta}(\zeta),
	\end{equation}
	with  links potential energy $P_{\ell}(q_{\ell})$, motors potential energy $P_{m g}(q_m)$ and the (coupling) potential energy due to the joints stiffness $P_{\zeta}(\zeta)$. The corresponding potential energy for linear springs is 
	\begin{equation}
	P_{\zeta}(\zeta)=\frac{1}{2}\zeta^{\top}K\zeta,
	\label{label:potential-flexible}
	\end{equation}
	with $\zeta:=q_m-q_{\ell}$   and the stiffness coefficients matrix $K\in\mathds{R}^{n\times n}$ is symmetric and positive definitive. Since $\text{rank}(B(q))=n_m$, the input matrix is given as $B(q)=[0_{n_{\ell}} B_m^{\top}(q_{m})]^{\top}$. Substitution of the above specifications in the Hamiltonian function  \eqref{eq:phmechanicalHamiltonian} and  the pH mechanical system \eqref{eq:phmechanical} results in the port-Hamiltonian model for a FJR explicitly given by
	\begin{equation}
	\begin{split}
	\begin{bmatrix}
	\dot q_{\ell}\\
	\dot q_m\\
	\dot p_{\ell}\\
	\dot p_m
	\end{bmatrix} &= \begin{bmatrix}
	0_{n_{\ell}} & 0_{n_m} & I_{n_{\ell}} & 0_{n_m} \\
	0_{n_{\ell}} & 0_{n_m}  & 0_{n_{\ell}} & I_{n_m} \\
	-I_{n_{\ell}} & 0_{n_m}  & -D_{\ell}(q_{\ell},p_{\ell}) & 0_{n_m} \\
	0_{n_{\ell}} & -I_{n_m}  & 0_{n_{\ell}} & -D_m(q_m,p_m)
	\end{bmatrix}\begin{bmatrix}
	\frac{\partial H}{\partial q_{\ell}}(q,p)\\
	\frac{\partial H}{\partial q_m}(q,p)\\
	\frac{\partial H}{\partial p_{\ell}}(q,p)\\
	\frac{\partial H}{\partial p_m}(q,p)
	\end{bmatrix} + \begin{bmatrix}
	0_{n_{\ell}}\\
	0_{n_m} \\
	0_{n_{\ell}}\\
	B_m(q_m)
	\end{bmatrix}u_m,\\
	y&=B_m^{\top}(q_m)\frac{\partial H}{\partial p_m}(q,p),
	\end{split}
	\label{eq:phmechanical-flexible}
	\end{equation}
	where $p_{\ell}=M_{\ell}(q_{\ell})\dot{q}_{\ell}$ and $p_m=M_m(q_m)\dot{q}_m$ are the  links and motors momenta, respectively; and $p=[p^{\top}_{\ell},p^{\top}_m]^{\top}$. Without loss of generality we take $B_m(q_m)=I_{n_m}$. The pH-FJR  \eqref{eq:phmechanical-flexible} can be rewritten  as the alternative model \eqref{eq:phmechanical-alternative} with
	\begin{equation}
	E(x)=\begin{bmatrix}
	S_{\ell}(q_{\ell},\dot{q}_{\ell}) -\frac{1}{2}\dot{M}_{\ell}(q_{\ell})& 0_{2n_m}\\
	0_{2n_{\ell}} & S_{m}(q_{m},\dot{q}_{m})-\frac{1}{2}\dot{M}_m(q_m) 
	\end{bmatrix}_{\dot{q}=M^{-1}(q)p},
	\label{eq:workless-flexible}
	\end{equation}
	with $S_{\ell}^{\top}(q_{\ell},p_{\ell})=-S_{\ell}(q_{\ell},p_{\ell})$ and $S_{m}^{\top}(q_{m},p_{m})=-S_{m}(q_{m},p_{m})$. We will also  denote the state of \eqref{eq:phmechanical-flexible} by $x:=[q^{\top},p^{\top}]^{\top}\in T^*\mathcal{Q}$.

	\subsection{Trajectory tracking control problem for FJRs}
	
	\subsubsection{Trajectory tracking problem:}
	Given a smooth reference trajectory  $q_{\ell d}(t)$ for the link's position $q_{\ell}(t)$, to design the input $u$ for the pH-FJR \eqref{eq:phmechanical-flexible} such that the link's position $q_{\ell}(t)$ converges asymptotically/exponentially to the reference trajectory $q_{\ell d}(t)$,  as $t\rightarrow \infty$  and all closed-loop system's trajectories are bounded.\\

	\subsubsection{Proposed solution:}
	Using the v-CBC method in Section \ref{subsection:v-dPBC}, design a control scheme with the following structure:
	\begin{equation}
	\zeta(x_v,x,t):=u_{v}^{ff}(x_v,x,t)+u_{v}^{fb}(x_v,x,t)
	\label{eq:control-tracking}
	\end{equation}
	where the \emph{feedforward-like} term $u_{v}^{ff}$ ensures that the closed-loop virtual  system has    the desired  trajectory $x_d(t)$ as steady-state solution, and the \emph{feedback} action $u_{v}^{fb}$ enforces  the closed-loop virtual system  to be differentially passive.
	
	\section{Trajectory-tracking control design and convergence analysis}
	Before presenting our main  contribution, we recall a v-CBC scheme  for a fully actuated rigid robot manipulators\cite{RodoAutomatica2017}  with $n_{\ell}$-dof,  which will be used in the main result. To this end, we assume that   this rigid robot is modeled as the pH system \eqref{eq:Hamiltonian2}, describing the links dynamics only. In order to avoid notation inconsistency between the rigid and flexible controllers, this is stressed   by adding the subscript $\ell$ to its  state and parameters in \eqref{eq:Hamiltonian2}, i.e., $x_{\ell}=[q_{\ell}^{\top},p_{\ell}^{\top}], D_{\ell}(x_{\ell}),E_{\ell}(x_{\ell}),B_{\ell}(q_{\ell})$ and $u_{\ell}$, respectively.
	
	\begin{lemma}[Reyes-B\'aez \cite{reyes2017virtual}]\label{lemma:fullstatecontroller-rigid}
		Consider the links dynamics given by \eqref{eq:Hamiltonian2} and its associated  virtual system \eqref{eq:phmechanicalvirtual}. Suppose that $\text{rank }B_{\ell}(q_{\ell})=n_{\ell}$ and let ${x}_{\ell d}=[q_{\ell d}^{\top},p_{\ell d}^{\top}]^{\top}$ be a smooth reference trajectory. Let us introduce the following error coordinates 
		\begin{equation}
		\tilde{{x}}_{\ell v}:=\begin{bmatrix}
		\tilde{{q}}_{\ell v}\\
		{\sigma_{\ell v}}
		\end{bmatrix}=\begin{bmatrix}
		{q}_{\ell v}-{q}_{\ell d} \hfill \\
		{p}_{\ell v}-{p}_{\ell r} 
		\end{bmatrix},
		\label{eq:changeofcoordinatesError}
		\end{equation}
		where the \emph{auxiliary momentum reference} $p_{\ell r}$ is given by
		\begin{equation}
		p_{\ell r}(\tilde{q}_{\ell v},t):=M(q)(\dot{q}_d-\phi_{\ell}(\tilde{q}_{\ell v})+\overline{v}_{\ell r}),
		\label{eq:auxiliarreference}
		\end{equation}
		with\footnote{The term $\overline{v}_{\ell r}$ is written explicitly  in \eqref{eq:changeofcoordinatesError} just for sake of clarity in the following developments. }  $\overline{v}_{\ell r}=0_{n_{\ell}}$, function $\phi_{\ell}:\mathcal{Q}_{\ell}\rightarrow T_{q_{\ell v}}\mathcal{Q}_{\ell}$ is such that $\phi_{\ell}(0_n)=0_n$; and $\Pi_{\ell}:\mathcal{Q}_{\ell}\times \mathds{R}_{\geq 0}\rightarrow \mathds{R}^{n_{\ell}\times n_{\ell}}$   a positive definite Riemannian metric tensor  satisfying  the inequality
		\begin{equation}
		\begin{split}
		\dot{\Pi}_{\ell}&(\tilde{q}_{\ell v},t)-\Pi_{\ell}(\tilde{q}_{\ell v},t)\frac{\partial \phi_{\ell}}{\partial \tilde{q}_{\ell v}}(\tilde{q}_{\ell v})-\frac{\partial \phi^{\top}_{\ell}}{\partial \tilde{q}_{\ell v}}(\tilde{q}_{\ell v})\Pi_{\ell}(\tilde{q}_{\ell v},t)\leq  -2\beta_{\ell}(\tilde{q}_{\ell v},t)  \Pi_{\ell}(\tilde{q}_{\ell v},t),
		\end{split}
		\label{eq:controllaw-positionmetricinquality}
		\end{equation}							
		with $\beta_{\ell}(\tilde{q}_{\ell v},t)>0$, uniformly.  Consider that   the $x_{\ell}$-parametrized composite control law given by
		\begin{equation}
		{u}_{\ell v}(x_{\ell v},x_{\ell},t):={u}_{\ell v}^{ff}(x_{\ell v},x_{\ell},t)+{u}_{\ell v}^{fb}(x_{\ell v},x_{\ell},t),
		\label{eq:controlaw}
		\end{equation}
		with
		\begin{equation}
		\begin{split}
		{u}_{\ell v}^{ff}&=\dot{p}_{\ell r}+\frac{\partial P_{\ell}}{\partial q_{\ell v}}(q_{\ell v})+\big[E_{\ell}(x_{\ell})+D_{\ell}(x_{\ell})\big] M^{-1}_{\ell}(q_{\ell})p_{\ell r},\quad
		{u}_{\ell v}^{fb}=-\int_{0}^{\tilde{{q}_{\ell v}}}\Pi_{\ell }(\xi_{\ell},t)d\xi_{\ell v}-{K}_{\ell d}M^{-1}_{\ell}(q_{\ell})\sigma_{\ell v}+\omega_{\ell},
		\end{split}
		\label{eq:controlaw1}
		\end{equation}
		where  the $i$-th row of  ${\Pi}_{\ell}(\tilde{q}_{\ell v},t)$  is a conservative vector field\footnote{This   ensures that the integral  in  \eqref{eq:controlaw1} is well defined and independent of the path connecting  $0$ and  $\tilde{q}_{\ell v}$.}, ${K}_{\ell d}>0$ and $\omega_{\ell}$ is an external input. Then, system \eqref{eq:phmechanicalvirtual} in closed-loop with \eqref{eq:controlaw}  is strictly differentially passive from  $\delta \omega_{\ell}$ to $\delta \overline{y}_{\sigma_{\ell v}}=M^{-1}_{\ell}(q_{\ell})\delta\sigma_{\ell v}$, with differential storage function given by
		\begin{equation}
		W_{\ell}(\tilde{x}_{\ell v},\delta \tilde{x}_{\ell v},t)=\frac{1}{2}\delta \tilde{x}^{\top}_{\ell v}\begin{bmatrix}
		\Pi_{\ell}(\tilde{q}_{\ell v},t)& 0_{n_{\ell}}\\
		0_{n_{\ell}} & M^{-1}_{\ell}(q_{\ell})
		\end{bmatrix}\delta \tilde{x}_{\ell v}.
		\label{eq:design-dLCF-sigma}
		\end{equation}					 
	\end{lemma}

	\subsection{Controller design for pH-FJRs}\label{section:controller-design-vdPBC}	
	Based on the v-CBC methodoloty described in Section \ref{subsection:v-dPBC}, the control scheme will be  designed as follows.
	
	\subsubsection{Step 1: Virtual mechanical system for a pH-FJR}
	Using \eqref{eq:workless-flexible}, the corresponding  virtual system \eqref{eq:phmechanicalvirtual} for the pH-FJR \eqref{eq:phmechanical-flexible} is  given by 
	\begin{equation}
	\begin{split}
	\dot{x}_v &=\begin{bmatrix}
	0_{n_\ell} & 0_{n_m} & I_{n_\ell} & 0_{n_m} \\
	0_{n_\ell} & 0_{n_m} & 0_{n_\ell} & I_{n_m} \\
	-I_{n_\ell} & 0_{n_m}  & -{\scriptsize (E_{\ell}(x_{\ell})+D_{\ell}(x_\ell ))}& 0_{n_m} \\
	0_{n_\ell} & -I_{n_m}  & 0_{n_\ell} & -{\scriptsize(E_{m}(x_m)+D_m(x_m) )}
	\end{bmatrix}\frac{\partial H_v}{\partial x_v}(x_v,x)+ \begin{bmatrix}
	0_{n_{\ell}}\\
	0_{n_m} \\
	0_{n_{\ell}}\\
	I_{n_m}
	\end{bmatrix}u_{mv},\\
	y_v&= \begin{bmatrix}
	0_{n_{\ell}} & 0_{n_m} &0_{n_{\ell}} &I_{n_m}
	\end{bmatrix}^{\top}\frac{\partial H_v}{\partial x_v}(x_v,x).
	\end{split}
	\label{eq:phmechanical-flexible-virtual}
	\end{equation}	
	with $H_v(x_v,x)$ as in \eqref{eq:phmechanicalHamiltonianvirtual} with respect to \eqref{label:kinetic-flexible}-\eqref{label:potential-flexible} and  $x_v=[q_{v}^{\top},p_{v}^{\top}]^{\top}\in T^*\mathcal{Q}$, with $q_v=[q_{\ell v}^{\top},q_{mv}^{\top}]^{\top}$ and $p_v=[p_{\ell v}^{\top},p_{mv}^{\top}]^{\top}$.

	\subsubsection{Step 2: Virtual differential passivity based controller design}
	Notice that in the links momentum dynamics of the virtual system  \eqref{eq:phmechanical-flexible-virtual}, that is, in 
	\begin{equation*}
	\dot{p}_{\ell v}=-\frac{\partial P_{\ell v}}{\partial q_{\ell v}}(q_{\ell})-\left[E_{\ell}(x_{\ell})+D_{\ell }(x_{\ell})\right]M^{-1}_{\ell}(q_{\ell})p_{\ell v}+K\zeta_v,
	\end{equation*}
	the potential force  $K\zeta_v=K(q_{mv}-q_{\ell v})$ acts in all the dof since $\text{rank}(K)=n_{\ell}$. Following the ideas  in \cite{brogliato1995global,ott2008passivity} of the passivity  approach, we want to find a desired motors position reference $q_{md}$ such that the  torque supplied by the springs  makes the position of the links   to track a desired   reference $q_{\ell d}(t)$. To this end, it is sufficient if the following potential forces relation   holds:
	\begin{equation}
	\frac{\partial P_{\zeta v}}{\partial q_{mv}}(q_{\ell v},q_{mv})= K(q_{mv}-q_{\ell v})=\frac{\partial \overline{P}_{\zeta v}}{\partial q_{mv}}(q_m,q_{md},q_{\ell v},t):=K(q_m-q_{md}) + u_{\ell v},
	\label{eq:PotentialEnergy-Matching}
	\end{equation}		
	for any $q_{mv}$ and $q_{\ell v}$, where $u_{\ell v}$ is an artificial input for the links dynamics, $P_{\zeta_v}(\zeta_v)$ is the virtual potential energy following the form in \eqref{label:potential-flexible} and $\overline{P}_{\zeta_v}(\zeta_v)$ is the target virtual  potential energy. The matching condition \eqref{eq:PotentialEnergy-Matching}  holds for  $q_{m d}=q_{\ell v}+K^{-1}u_{\ell v}$.

	\begin{proposition}\label{proposition:fullstatecontroller-flexible}
		Consider the original system \eqref{eq:phmechanical-flexible} and its virtual system   \eqref{eq:phmechanical-flexible-virtual}.  Consider also  the controller $u_{\ell v}$ in  \eqref{eq:controlaw1}. Let  ${x}_{m d}=[q_{m d}^{\top},p_{m d}^{\top}]^{\top}$ be the  motor  reference state, with $q_{m d}=q_{\ell v}+K^{-1}u_{\ell v}$. Let us introduce the motors error coordinates as
		\begin{equation}
		\tilde{{x}}_{mv}:=\begin{bmatrix}
		\tilde{{q}}_{mv}\\
		{\sigma_{mv}}
		\end{bmatrix}=\begin{bmatrix}
		{q}_{mv}-{q}_{md}\\
		{p}_{mv}-{p}_{mr}
		\end{bmatrix},
		\label{eq:changeofcoordinatesError-motor}
		\end{equation}
		where the artificial motor momentum reference $p_{mr}$ is defined by
		\begin{equation}
		p_{mr}:=M_{m}(q_m)\big(\dot{q}_{md}-\phi_{m}(\tilde{q}_{mv})+\overline{v}_{mr}\big),
		\label{eq:auxiliarreference-motor}
		\end{equation}
		with $\delta\overline{v}_{mr}=-\Pi^{-1}_{m}(\tilde{q}_{m v},t) K^{\top}M_{\ell}^{-\top}(q_{\ell}){\sigma_{\ell v}} $, function $\phi_{m}:\mathcal{Q}_m\rightarrow T_{\tilde{q}_{m v}}\mathcal{Q}_m$  and a positive definite Riemannian metric $\Pi_{m}:\mathcal{Q}_m\times \mathds{R}_{\geq 0}\rightarrow \mathds{R}^{n_m\times n_m}$  satisfying  the inequality
		\begin{equation}
		\begin{split}
		\dot{\Pi}_{m}(\tilde{q}_{m v},t)-&\Pi_{m }(\tilde{q}_{m v},t)\frac{\partial \phi_{m}}{\partial \tilde{q}_{m v}}(\tilde{q}_{m v})-\frac{\partial \phi_{m}^{\top}}{\partial \tilde{q}_{m v}}(\tilde{q}_{m v}) \Pi_{m}(\tilde{q}_{m v},t)\leq  -2\beta_{m}(\tilde{q}_{m v},t) \Pi_{m}(\tilde{q}_{m v},t),
		\end{split}
		\label{eq:controllaw-positionmetricinquality-motor}
		\end{equation}							
		with $\beta_{m}(\tilde{q}_{m v},t)>0$, uniformly. Assume that the $i$-th  row of $\Pi_{m}(\tilde{q}_{m v},t)$ is a conservative vector field\footnote{This   ensures that the integral  in  \eqref{eq:controlaw1} is well defined and independent of the path connecting  $0$ and  $\tilde{q}_{mv}$.}. Then, the virtual system \eqref{eq:phmechanical-flexible-virtual} in closed-loop with the  control law given by  
		\begin{equation}
		{u}_{mv}(x_{v},x,t):={u}_{mv}^{ff}(x_{v},x,t)+{u}_{mv}^{fb}(x_{v},x,t), 
		\label{eq:controlaw-motor}
		\end{equation} 
		with
		\begin{equation}
		\begin{split}
		{u}_{mv}^{ff}(x_{v},x,t)&=\dot{p}_{mr}+\frac{\partial P_{m}}{\partial q_{m v}}(q_{m v})+k\zeta_v +\big[E_m(x_m)+D_m(x_m)\big] M_m^{-1}(q_m)p_{mr},\\
		{u}_{mv}^{fb}(x_{v},x,t)&=-\int_{0_{n_m}}^{\tilde{{q}}_{mv}}\Pi_{m}(\xi_{m v},t)\text{d}\xi_{mv}-{K}_{md}M^{-1}_m(q_m)\sigma_{mv}+\omega_m,
		\end{split}
		\label{eq:controlaw-motor1}
		\end{equation}
		is strictly differentially passive  from $\delta \omega$ to $\delta y_{\sigma_{v}}=M^{-1}(q)\delta\sigma_{v}$  with respect to the   differential storage function
		\begin{equation}
		W(\tilde{x}_v,\delta \tilde{x}_v,t)=\frac{1}{2}\delta \tilde{x}_v^{\top}\begin{bmatrix}
		\Pi_{\tilde{q}_v}(\tilde{q}_v,t)& 0_n\\
		0_n & M^{-1}(q)
		\end{bmatrix}\delta \tilde{x}_v,
		\label{eq:design-dLCF-sigma-flexible}
		\end{equation}		
		where the error coordinate is $\tilde{x}_v=[\tilde{q}_v^{\top},\sigma_v^{\top}]^{\top}$, with $\tilde{q}_v:=[\tilde{q}_{\ell v}^{\top},\tilde{q}_{mv}^{\top}]^{\top}$   and $\sigma_v:=[\sigma_{\ell v}^{\top},\sigma_{mv}^{\top}]^{\top}$. Matrix ${K}_{md}>0$ is  a constant derivative gain, $ \omega=[\omega_{\ell}^{\top},\omega_m^{\top}]^{\top} $ is an external input   and  ${\Pi}_{\tilde{q}_v}(\tilde{q}_v,t):=\text{diag}\{\Pi_{\ell}(\tilde{q}_{\ell v},t),\Pi_m(\tilde{q}_{mv},t)\}$. Moreover,   \eqref{eq:design-dLCF-sigma-flexible} qualifies as differential Lyapunov function and the virtual system \eqref{eq:phmechanical-flexible-virtual} in closed-loop with the  control law \eqref{eq:controlaw-motor} is contractive for $\omega=0_n$.
	\end{proposition}
	\subsubsection{Step 3: Trajectory tracking controller for the pH-FJR}			
	Notice that by construction,   the origin $(\tilde{q}_v,\sigma_v)=(0_n,0_n)$ is a solution of the closed-loop system if $\omega=0_n$.  Using this fact, in the next result we propose a family of trajectory-tracking controllers for the pH-FJR \eqref{eq:phmechanical-flexible}.
	
	\begin{corollary}\label{corollary:Actualtrackingcontrol}
		Consider    the virtual controller \eqref{eq:controlaw-motor} and let $q_{\ell d}(t)\in\mathcal{Q}_{\ell}$ be a reference  time-varying trajectory.  Suppose that the flexible joints robot   \eqref{eq:phmechanical-flexible} is controlled by the scheme
		\begin{equation}
		u_{m}(x,t):=u_{mv}(x,x,t).
		\label{eq:control-flexible}
		\end{equation}
		Then, the  links position $q_{\ell}$ of the  closed-loop system converges globally and exponentially   to the trajectory $q_{ d}(t)$,  with rate 
		\begin{equation}
		\begin{split}
		\beta&=2\min\{\beta_{\tilde{q}}(\tilde{q}_v,t),\lambda_{\min}\{D(x)+K_d\}\lambda_{\min}\{M^{-1}(q)\}\}.	
		\end{split}
		\label{eq:position-convergencerate}
		\end{equation}
	\end{corollary}		 		
	\subsection{Properties of the closed-loop virtual system}\label{section:closed-loop-properties}
	\subsubsection{Structural properties}
	In the following result we show that system \eqref{eq:phmechanical-flexible-virtual}  in closed-loop with \eqref{eq:controlaw-motor} preserves the structure of  the variational dynamics  \eqref{eq:feedbackinterconnectionsystemsvariational}.
	\begin{corollary}\label{corollary:strucutr-preserving}
		Consider system  \eqref{eq:phmechanical-flexible-virtual}  in closed-loop with \eqref{eq:controlaw-motor}.  Then the closed-loop variational dynamics satisfies Lemma \ref{lemma:dPBC}, in coordinates $\tilde{x}_v$, with 
		\begin{equation}
		\begin{split}
		\Pi(\tilde{x}_v,t)&=\begin{bmatrix}
		\Pi_{\ell}(\tilde{q}_{\ell v},t) & 0_{n_{m}} &	0_{n_{\ell}}   & 0_{n_m}\\
		0_{n_{\ell}} & \Pi_{m}(\tilde{q}_{\ell v},t) & 0_{n_{\ell}} &  0_{n_{m}}\\
		0_{n_{\ell}} & 0_{n_{m}} & M^{-1}_{\ell }(q_{\ell}) &  0_{n_{m}}\\
		0_{n_{\ell}}  &   0_{n_m} & 0_{n_{\ell}} &  M^{-1}_{m }(q_{m})
		\end{bmatrix}; \Xi(\tilde{x}_v,t)=\begin{bmatrix}
		0_{n_{\ell}} & 0_{n_{m}} & I_{n_{\ell}} & 0_{n_m}\\
		0_{n_{\ell}} & 0_{n_{m}} & -\Pi_m^{-1}(\tilde{q}_{mv},t)K^{\top} &  I_{n_{m}}\\
		-I_{n_{\ell}} & K\Pi_m^{-1}(\tilde{q}_{mv},t) & -S_{\ell}(q_{\ell},p_{\ell}) &  0_{n_{m}}\\
		0_{n_{\ell}}  &  - I_{n_{m}} & 0_{n_{\ell}} &  -S_{m}(q_{m},p_{m})
		\end{bmatrix};\\
		\Upsilon(\tilde{x}_v,t)&=\begin{bmatrix}
		\frac{\partial \phi_{\ell}}{\partial \tilde{q}_{\ell v}}{\Pi}^{-1}_{\ell }(\tilde{q}_{\ell v},t) & 0_{n_{m}} &	0_{n_{\ell}}   & 0_{n_m}\\
		0_{n_{\ell}} & \frac{\partial \phi_{m}}{\partial \tilde{q}_{m v}}{\Pi}^{-1}_{m }(\tilde{q}_{m v},t) & 0_{n_{\ell}} &  0_{n_{m}}\\
		0_{n_{\ell}} & 0_{n_{m}} & \left(D_{\ell}+K_{\ell d}-\frac{1}{2}\dot{M}_{\ell}(q_{\ell})\right) &  0_{n_{m}}\\
		0_{n_{\ell}}  &   0_{n_m} & 0_{n_{\ell}} & \left(D_{m}+K_{m d}-\frac{1}{2}\dot{M}_{m}(q_{m})\right)
		\end{bmatrix}; 
		\Psi=\begin{bmatrix}
		0_{n_{\ell}} & 0_{n_m}\\
		0_{n_{\ell}} & 0_{n_m}\\
		I_{n_{\ell}}& 0_{n_m}\\
		0_{n_{\ell}} & I_{n_m}\\			  
		\end{bmatrix}, 			
		\end{split}
		\label{eq:feedbackinterconnectionsystemsvariational-flexible}
		\end{equation}
		and   $\Theta(x_v,t)$   given by the Jacobian  of  $\tilde{x}_v=x_v-x_d(x_v,t)$, with respect to $x_v$, where  desired state $x_d:=[q_{\ell d}^{\top}, q_{m d}^{\top}, p_{\ell r}^{\top},p_{m r}^{\top}]^{\top}$.
	\end{corollary}
	In other words, the statement in Corollary \ref{corollary:strucutr-preserving} tells us that the differential transformation $\Theta(x_v,t)$ is implicitly constructed via the design procedure of Proposition \ref{proposition:fullstatecontroller-flexible}. Furthermore,  notice that  the closed-loop  dynamics of both,   $\sigma_{\ell v}$ and   $\sigma_{mv}$ in \eqref{eq:design-closed-complete} are actuated by $\omega_{\ell}$ and $\omega_m$, respectively. This is in fact a direct consequence of the  potential energy matching condition \eqref{eq:PotentialEnergy-Matching},  making possible to rewrite the error dynamics as a "fully-actuated" system    in \eqref{eq:momentum-error-dynamics-Kz}.     Such   interpretation of the closed-loop system \eqref{eq:design-closed-complete} allows us to extend some of the structural properties of the v-CBC scheme for fully-actuated systems in our previous work Reyes-B\'aez \cite{RodoAutomatica2017}.
	\begin{corollary}\label{corollary:preserving-variational-pH-structure}
		Consider system  \eqref{eq:phmechanical-flexible-virtual}  in closed-loop with \eqref{eq:controlaw-motor}.  Assume   that the Jacobian matrices $\frac{\partial \phi_{\ell}}{\partial \tilde{q}_{\ell}}(\tilde{q}_{\ell v})$ and  $\frac{\partial \phi_{m}}{\partial \tilde{q}_{m}}(\tilde{q}_{m v})$	are symmetric and assume that the  products $\Pi_{\ell}(\tilde{q}_{\ell v},t)\frac{\partial \phi_{\ell}}{\partial \tilde{q}_{\ell}}(\tilde{q}_{\ell v})$ and  $\Pi_{m}(\tilde{q}_{m v},t)\frac{\partial \phi_{m}}{\partial \tilde{q}_{m}}(\tilde{q}_{m v})$ commute. Then  the closed-loop variational system preserves  the structure of the variational   pH-like system \eqref{eq:variationalvirtualpHsystem}, in coordinates $\tilde{x}_v$, with  
		\begin{equation}
		\frac{\partial^2 \tilde{H}_v}{\partial x_v^2}(\tilde{x}_v,x)=\Pi(\tilde{x}_v,t) \quad \tilde{J}_v(\tilde{x}_v,t)=\Xi(\tilde{x}_v,t),\quad\tilde{R}_v(\tilde{x}_v,t)=\Upsilon(\tilde{x}_v,t),\quad \tilde{g}:=\Psi^{\top}.
		\label{eq:preserving-variational-pH-structure}
		\end{equation}
	\end{corollary}
	Notice that all   matrices in \eqref{eq:preserving-variational-pH-structure}  that define  the variational system in Corollary \ref{corollary:preserving-variational-pH-structure} are  state and time dependent, while  the ones of the variational system \eqref{eq:variationalvirtualpHsystem} are only time dependent; in this sense the system in Corollary \ref{corollary:preserving-variational-pH-structure}  is more general. However, despite of the structure of the variational dynamics \eqref{eq:variationalvirtualpHsystem} is preserved, the system defined by \eqref{eq:preserving-variational-pH-structure} does not necessarily correspond to  a  pH-like mechanical system as in \eqref{eq:pHlike-virtual}. This would be the case under the   following if and only if conditions:
	\begin{equation}
	\Pi_{\ell }(\tilde{q}_{\ell  v},t)= \frac{\partial \phi_{\ell}}{\partial \tilde{q}_{\ell v}}(\tilde{q}_{\ell v})\quad\text{and}\quad \Pi_{m}(\tilde{q}_{m  v},t)= \frac{\partial \phi_{m}}{\partial \tilde{q}_{m v}}(\tilde{q}_{m v})=\Lambda_{m}
	\label{eq:pH-like-structure-preserving-condition}
	\end{equation}
	where $\Lambda_{m}$ is a constant symmetric and positive definite matrix. Indeed, substitution   in the closed-loop system \eqref{eq:design-closed-complete} gives
	\begin{equation}
	\begin{split}
	\dot{\tilde{x}}_v&= {\begin{bmatrix}
		-I_{n_{\ell}}  & 0_{n_{m}} & I_{n_{\ell}} & 0_{n_m}\\
		0_{n_{\ell}} & -I_{n_{m}} & -\Lambda_m^{-1}K^{\top} &  I_{n_{m}}\\
		-I_{n_{\ell}}  & K\Lambda_m^{-1} & -\left(E_{\ell}(x_{\ell})+D_{\ell}(x_{\ell})+K_{\ell d}\right) &  0_{n_{m}}\\
		0_{n_{\ell}}  & - I_{n_{m}}  & 0_{n_{\ell}} &  -\left(E_{m}(x_m)+D_{m}(q_m)+K_{m d}\right)
		\end{bmatrix}}\frac{\partial \tilde{H}_v}{\partial \tilde{x}_{ v}}(\tilde{x}_v,x)+\begin{bmatrix}
	0_{n_{\ell}} & 0_{n_m}\\
	0_{n_{\ell}} & 0_{n_m}\\
	I_{n_{\ell}}& 0_{n_m}\\
	0_{n_{\ell}} & I_{n_m}\\			  
	\end{bmatrix}\omega.\\
	\tilde{	y}_v&=\begin{bmatrix}
	0_{n_{\ell}} &  	0_{n_{\ell}} & I_{n_{\ell}} & 0_{n_{\ell}}\\
	0_{n_m} & 0_{n_m}&   0_{n_m} &   I_{n_m}\\			  
	\end{bmatrix}\frac{\partial \tilde{H}_v}{\partial \tilde{x}_{ v}}(\tilde{x}_v,x)
	\end{split}
	\label{eq:design-closed-complete-structure-Lamnda}
	\end{equation}      		
	where the $x$-parametrized closed-loop error Hamiltonian function  is	given by
	\begin{equation}
	\tilde{H}_v(\tilde{x}_v,{x})=\frac{1}{2}\tilde{x}_v^{\top}\Pi(x)\tilde{x}_v=\int_{0_{n_{\ell}}}^{\tilde{q}_{\ell v}}\phi_{\ell}(\overline{q}_{\ell v})d\overline{q}_{\ell v}+\frac{1}{2}\tilde{q}_{m v}^{\top}\Lambda_m \tilde{q}_{m  v} +\frac{1}{2} \sigma_{  v}^{\top}M^{-1}(q)\sigma_{ v}.
	\label{eq:closed-looperror-Hamiltonianfunction}
	\end{equation}			
	

\subsubsection{Differential passivity properties}
	In this part we   give a differential passivity interpretation of 	system  \eqref{eq:phmechanical-flexible-virtual}  in closed-loop with the   scheme \eqref{eq:controlaw-motor}.  Before stating the result, let us write the closed-loop variational system for the links  error state $\tilde{x}_{\ell v}$ as\footnote{ For  sake of presentation, we  explicitly consider the two components of vector $\overline{v}_r=[\overline{v}_{\ell r}^{\top},\overline{v}_{mr}^{\top}]^{\top}$   in \eqref{eq:variables-proof}, even though we know in advance that $\overline{v}_{\ell r}=0_{n_{\ell}}$.} 
	\begin{equation}
	\begin{split}
	\begin{bmatrix}
	\delta\dot{\tilde{q}}_{\ell v}\\			
	\delta\dot{\sigma}_{\ell v}\\
	\end{bmatrix}&= {\begin{bmatrix}
		-\frac{\partial \phi_{\ell}}{\partial \tilde{q}_{\ell v}}(\tilde{q}_{\ell v}){\Pi}^{-1}_{\ell } (\tilde{q}_{\ell v},t) & I_{n_{\ell}} \\
		-I_{n_{\ell}} & -\left(E_{\ell}(x_{\ell})+D_{\ell}(x_{\ell})+K_{\ell d}\right)\\
		\end{bmatrix}} \begin{bmatrix}
	{\Pi}_{\ell }(\tilde{q}_{\ell},t)\delta\tilde{q}_{\ell v}\\			
	M^{-1}_{\ell}(q_{\ell})\delta\sigma_{\ell v}\\
	\end{bmatrix}+\begin{bmatrix}		
	I_{n_{\ell}} & 0_{n_{\ell}}\\
	0_{n_{\ell}} & I_{n_{\ell}}\\			  
	\end{bmatrix}\begin{bmatrix}
	\delta\overline{v}_{\ell r}\\
	K\delta\tilde{q}_{md}+\delta\omega_{\ell} 
	\end{bmatrix}
	\end{split}
	\label{eq:variational-link-dynamics}
	\end{equation}
	which by Lemma \ref{corollary:strucutr-preserving}, preserves the structure of  \eqref{eq:feedbackinterconnectionsystemsvariational} and is given by 
	\begin{equation}
	\begin{split}
	\delta\tilde{x}_{\ell v}&= \underbrace{{\begin{bmatrix}
			-\frac{\partial \phi_{\ell}}{\partial \tilde{q}_{\ell v}}(\tilde{q}_{\ell v}){\Pi}^{-1}_{\ell } (\tilde{q}_{\ell v},t) & I_{n_{\ell}} \\
			-I_{n_{\ell}} & -\left(E_{\ell}(x_{\ell})+D_{\ell}(x_{\ell})+K_{\ell d}\right)\\
			\end{bmatrix}}}_{\Xi_{\ell}(\tilde{x}_{\ell},t)-\Upsilon_{\ell}(\tilde{x}_{\ell},t)}\frac{\partial^2 \tilde{H}_{\ell}}{\partial \tilde{x}_{\ell v}^{2}}(\tilde{x}_{\ell v},x_{\ell},t)\delta \tilde{x}_{\ell v}+\begin{bmatrix}		
	I_{n_{\ell}} & 0_{n_{\ell}}\\
	0_{n_{\ell}} & I_{n_{\ell}}\\			  
	\end{bmatrix}\begin{bmatrix}
	\delta\overline{v}_{\ell r}\\
	\delta\overline{\omega}_{\ell}
	\end{bmatrix}\\
	\delta\tilde{y}_{\ell}&=\begin{bmatrix}		
	I_{n_{\ell}} & 0_{n_{\ell}}\\
	0_{n_{\ell}} & I_{n_{\ell}}\\			  
	\end{bmatrix}\frac{\partial^2 \tilde{H}_{\ell}}{\partial \tilde{x}_{\ell v}^{2}}(\tilde{x}_{\ell v},x,t)\delta \tilde{x}_{\ell v}
	\end{split}
	\label{eq:variational-link-dynamics1}
	\end{equation}	
	where $\delta\overline{\omega}_{\ell}=(K\delta\tilde{q}_{md}+\delta\omega_{\ell})$ and the Riemannian metric of \eqref{eq:feedbackinterconnectionsystemsvariational}, in this case, is given by the \emph{Hessian} of the   energy-like function
	\begin{equation}
	\tilde{H}_{\ell}(\tilde{x}_{\ell v},x_{\ell},t):=\frac{1}{2} \tilde{x}_{\ell v}^{\top}\begin{bmatrix}
	{\Pi}^{-1}_{\ell } (\tilde{q}_{\ell v},t)& 0_{n_{\ell}}\\
	0_{n_{\ell}} & M^{-1}_{\ell}(q_{\ell})
	\end{bmatrix}  \tilde{x}_{\ell v}.
	\end{equation}
	Moreover, the map $\begin{bmatrix}
	\delta\overline{v}_{\ell r}^{\top}&
	\delta\omega_{\ell}^{\top} 
	\end{bmatrix}^{\top}\mapsto \delta \tilde{y}_{\ell}$ is strictly differentially passive with respect to the differential storage function  
	\begin{equation}
	W_{\ell}(\tilde{x}_{\ell v},\delta\tilde{x}_{\ell v},t)=\frac{1}{2}\delta \tilde{x}_{\ell v}^{\top}\frac{\partial^2 \tilde{H}_{\ell}}{\partial \tilde{x}_{\ell v}^{2}}(\tilde{x}_{\ell v},x_{\ell},t)\delta \tilde{x}_{\ell v}.
	\label{eq:differentionalhamiltonianfunction-link}
	\end{equation}	
	Similarly,  the variational dynamics of the motor  error state $\tilde{x}_{m v}$   is 
	\begin{equation}
	\begin{split}
	\begin{bmatrix}
	\delta\dot{\tilde{q}}_{m v}\\			
	\delta\dot{\sigma}_{m v}\\
	\end{bmatrix}&= {\begin{bmatrix}
		-\frac{\partial \phi_{m}}{\partial \tilde{q}_{m v}}(\tilde{q}_{m v}){\Pi}^{-1}_{m } (\tilde{q}_{m v},t) & I_{n_{m}} \\
		-I_{n_{m}} & -\left(E_{m}(x_{m})+D_{m}(q_{m})+K_{m d}\right)\\
		\end{bmatrix}} \frac{\partial^2 \tilde{H}_{m}}{\partial \tilde{x}_{m v}^{2}}(\tilde{x}_{m v},x_m,t)\delta \tilde{x}_{m v}+\begin{bmatrix}		
	I_{n_{m}} & 0_{n_{m}}\\
	0_{n_{m}} & I_{n_{m}}\\			  
	\end{bmatrix}\begin{bmatrix}
	\delta\overline{v}_{m r}\\
	\delta\overline{\omega}_{m} 
	\end{bmatrix}\\
	\delta\tilde{y}_{m}&=\begin{bmatrix}		
	I_{n_{\ell}} & 0_{n_{\ell}}\\
	0_{n_{\ell}} & I_{n_{\ell}}\\			  
	\end{bmatrix}\frac{\partial^2 \tilde{H}_{m}}{\partial \tilde{x}_{m v}^{2}}(\tilde{x}_{m v},x_m,t)\delta \tilde{x}_{mv}	
	\end{split}
	\label{eq:variational-motor-dynamics}	
	\end{equation}	
	with $\delta\overline{v}_{m r}={\Pi}_{m }(\tilde{q}_{m},t)K^{\top}M^{-1}_{\ell}(q_{\ell})\delta\sigma_{\ell v}$, $\delta{\omega}_{m} =\delta\overline{\omega}_{m}$ and energy-like function 
	\begin{equation}
	\tilde{H}_{m}(\tilde{x}_{m v},x_m,t):=\frac{1}{2} \tilde{x}_{mv}^{\top}\begin{bmatrix}
	{\Pi}^{-1}_{m} (\tilde{q}_{m v},t)& 0_{n_m}\\
	0_{n_m} & M^{-1}_{m}(q_{m})
	\end{bmatrix}  \tilde{x}_{m v}.
	\end{equation}
	Also the map $\begin{bmatrix}
	\delta\overline{v}_{m r}^{\top}&
	\delta\omega_{m}^{\top} 
	\end{bmatrix}^{\top}\mapsto \delta \tilde{y}_{m}$ is strictly differentially passive with respect to the differential storage function  
	\begin{equation}
	W_{m}(\tilde{x}_{m v},\delta\tilde{x}_{mv},t)=\frac{1}{2}\delta \tilde{x}_{m v}^{\top}\frac{\partial^2 \tilde{H}_{m}}{\partial \tilde{x}_{m v}^{2}}(\tilde{x}_{m v},x_m,t)\delta \tilde{x}_{m v}.
	\label{eq:differentionalhamiltonianfunction-motor}
	\end{equation}	
	These show  that the corresponding closed-loop links and motor systems are differentially passive.

	\begin{corollary}\label{corollary:feedback-interconnection}
		Consider the closed-loop links and motors systems together with their variational dynamics in  \eqref{eq:variational-link-dynamics1}  and \eqref{eq:differentionalhamiltonianfunction-motor}, respectively.  Then, the resulting interconnected system  via the law
		\begin{equation}
		\begin{bmatrix}
		\delta\overline{v}_{\ell r}\\
		\delta\overline{\omega}_{\ell}\\
		\delta\overline{v}_{m r}\\
		\delta\overline{\omega}_{m} 		
		\end{bmatrix}=\begin{bmatrix}
		0_{n_{\ell}}& 0_{n_{\ell}} & 0_{n_{m}} & 0_{n_{m}}\\
		0_{n_{\ell}}& 0_{n_{\ell}} &  K\Pi_{m}(\tilde{q}_{m v},t) & 0_{n_{m}}\\
		0_{n_{\ell}}&-\Pi_{m}(\tilde{q}_{m v},t)K^{\top} & 0_{n_{m}} & 0_{n_{m}}\\
		0_{n_{\ell}}& 0_{n_{\ell}} & 0_{n_{m}} & 0_{n_{m}}\\
		\end{bmatrix}\begin{bmatrix}
		\delta \tilde{y}_{{\ell}v}\\ 
		\delta \tilde{y}_{{m}v}
		\end{bmatrix}+\begin{bmatrix}
		0_{n_{\ell}} & 0_{n_m}\\
		I_{n_{\ell}} & 0_{n_m}\\
		0_{n_{\ell}}& 0_{n_m}\\
		0_{n_{\ell}} & I_{n_m}\\			  
		\end{bmatrix}\delta \omega
		\label{eq:diff-passivity-interconnection}
		\end{equation}
		is differentially passive system with storage function $W(\tilde{x}_v,\delta \tilde{x}_v,t)=W_{\ell}(\tilde{x}_{\ell v},\delta\tilde{x}_{\ell v},t)+W_{m}(\tilde{x}_{m v},\delta\tilde{x}_{mv},t)$.		
	\end{corollary}

	The statement in Corollary \ref{corollary:feedback-interconnection} is closely related to the main result in the work of Jard\'on-Kojakhmetov \cite{hilde}, where a tracking controller for FJRs was developed using the singular perturbation approach. Under time-scale separation assumptions,  in that work it is shown that controller design can be performed in a \emph{composite manner} as $u=u_s+u_f$, where the links dynamics \emph{slow} controller $u_s$ and the motors dynamics  \emph{fast} controller $u_f$ can be designed separately. Both systems, the slow and fast, are fully actuated and standard control techniques for rigid robots can be applied as long as exponential stability can be guaranteed. 
	
	In this work \emph{we do not make any explicit assumption on time scale separation} in the design process. Nevertheless, due to  condition \eqref{eq:PotentialEnergy-Matching}, we require that the motors position error dynamics converges "faster" than the links one since $K\zeta_v=u_{\ell v}+K\tilde{q}_{mv}$. In this sense,  the singular perturbation approach can be used for adjusting the convergence rate of the closed-loop system.

	\subsubsection{Passivity properties}
	It is easy to verify that the map $\omega\mapsto \tilde{y}_v$ is  cyclo-passive with storage function \eqref{eq:closed-looperror-Hamiltonianfunction} for the closed-loop  system \eqref{eq:design-closed-complete-structure-Lamnda}; in fact strictly passive under conditions \eqref{eq:controllaw-positionmetricinquality} and \eqref{eq:controllaw-positionmetricinquality-motor}. This is a direct consequence of the pH-like structure preserving conditions  \eqref{eq:pH-like-structure-preserving-condition}. Furthermore, passivity of \eqref{eq:design-closed-complete-structure-Lamnda} is independent of the properties on $\phi_{\ell}(\tilde{q}_{\ell v})$ and $\Lambda_m$. Nevertheless, we have to be careful in  how we design $\Pi_{\ell}\phi_{\ell}(\tilde{q}_{\ell v})$ since passivity of system  \eqref{eq:design-closed-complete-structure-Lamnda} does not necessarily imply differential passivity; the converse is true.
	
	In what follows we give necessary and sufficient conditions  on  $\phi_{\ell}(\tilde{q}_{\ell v})$ and $\phi_{m}(\tilde{q}_{m v})=\Lambda_m\tilde{q}_{mv}$ in order to guarantee strict differential passivity and strict passivity of the closed-loop system \eqref{eq:design-closed-complete-structure-Lamnda} simultaneously. To this end, let us recall the following:	
	\begin{defin}[\cite{pavlov2006}]\label{definicion:incrementalPassivity-static}
		The map $\chi(z)$ is incrementally passive if it satisfies the following monotonicity condition: 	
		\begin{equation}
		\left[\chi(z_{2})-\chi(z_{1})\right]^{\top}(z_2-z_1)\geq 0,
		\label{eq:incrementalpassivity-static}
		\end{equation}
		for any $z_1$ and $z_2$. The property is strict if the inequality  \eqref{eq:incrementalpassivity-static} is strict.
	\end{defin}	
	\begin{lemma}[\cite{RodoAutomatica2017}]\label{lemma:incrementa-static}
		If $\Pi_{\ell}(\tilde{q}_{\ell v},t)$ and $\Pi_{m}(\tilde{q}_{m v},t)$ are constant in \eqref{eq:controllaw-positionmetricinquality} and \eqref{eq:controllaw-positionmetricinquality-motor}, respectively. Then, the maps $\chi_{\ell}(\tilde{q}_{\ell v})=\Pi_{\ell}\phi_{\ell}(\tilde{q}_{\ell v})$ and $\chi_{m}(\tilde{q}_{m v})=\Pi_{m}\Lambda_m\tilde{q}_{mv}$ are  strictly incrementally passive.  
	\end{lemma}
	As said before,  conditions in  Lemma \ref{lemma:incrementa-static} are only sufficient for the incremental stability property of the above maps. However, there may exist incrementally passive maps which do not satisfy inequalities \eqref{eq:controllaw-positionmetricinquality} and \eqref{eq:controllaw-positionmetricinquality-motor}. The following result gives  necessary and sufficient conditions to guarantee both properties, simultaneously.
	\begin{proposition}\label{proposition:iff-conditons-phi}
		Consider the maps $\chi_{\ell}(\tilde{q}_{\ell v})=\Pi_{\ell}\phi_{\ell}(\tilde{q}_{\ell v})$ and $\chi_{m}(\tilde{q}_{m v})=\Pi_{m}\Lambda_m\tilde{q}_{mv}$, with $\Pi_{\ell}$ and $\Pi_{m}$ symmetric positive definite and constant. Inequalities \eqref{eq:controllaw-positionmetricinquality} and \eqref{eq:controllaw-positionmetricinquality-motor} are satisfied if and only if the following condition  holds:
		\begin{equation}
		\begin{split}
		(\tilde{q}_{k v,2}-\tilde{q}_{k v,1})^{\top}\left[\chi_{k}(\tilde{q}_{k v,2})- \chi_{k}(\tilde{q}_{k v,1})\right]\geq 2\beta_{\tilde{q}_{k v}}(\tilde{q}_{k v,2}-\tilde{q}_{k v,1})^{\top}\Pi_{k }(\tilde{q}_{k v,2}-\tilde{q}_{k v,1})>0, \text{ for all } \tilde{q}_{k v,1},\tilde{q}_{k v,2}\quad \text{and for all } k\in\{\ell,m\}.
		\label{eq:necessarycondition-incrementalpassicity-link}
		\end{split}
		\end{equation} 
	\end{proposition}
	If conditions of Proposition \ref{proposition:iff-conditons-phi} are not satisfied,  using  Lemma \ref{lemma:incrementa-static} we still can find  (incrementally/shifted) passive maps $\chi_{\ell}$ and $\chi_m$ that make \eqref{eq:closed-looperror-Hamiltonianfunction} a Lyapunov function for  system \eqref{eq:design-closed-complete-structure-Lamnda} with minimum at the origin.  However, under Lemma \ref{lemma:incrementa-static}  it is not possible to ensure  that the unique steady-state trajectory of the closed-loop system  \eqref{eq:design-closed-complete-structure-Lamnda} is $x_d:=[q_{\ell d}^{\top}, q_{m d}^{\top}, p_{\ell r}^{\top},p_{m r}^{\top}]^{\top}$, because the contractivity conditions  \eqref{eq:controllaw-positionmetricinquality} and \eqref{eq:controllaw-positionmetricinquality-motor} are not necessarily satisfied.

	\section{Experiments evaluation of tracking controller for FJRs}
	In this section we present the design procedure and experimental evaluation of two schemes which lie in the family of v-CBC controllers as discussed in Section \ref{section:controller-design-vdPBC}. Each of these tracking controllers exhibits different closed-loop properties with respect to Section \ref{section:closed-loop-properties}.  Furthermore, by Corollary \ref{corollary:strucutr-preserving}, the closed-loop variational dynamics structure can be used as a \emph{qualitative tool} for gain tuning,   due to matrices in  \eqref{eq:feedbackinterconnectionsystemsvariational-flexible} allow us to have   a  \emph{clear physical interpretation} of the controller design parameters     \eqref{eq:controlaw-motor}, in terms of linear mass-spring-dampers systems which are modulated\footnote{These linear mass-spring-dampers systems have state $x_v$,  and are modulated by the "parameter" $x$ in the sense that their corresponding state space is given by $T_x\mathcal{X}$.} by the actual FJR's state $x$.  
	For short, considering the original state $\tilde{x}$, we  denote this family of controllers as
	\begin{equation*}
	\left(\Pi(\tilde{q},t), K_d, \phi(\tilde{q})\right)\text{-controller}.
	\end{equation*}

	For all  experiments we consider  $t\mapsto q_{\ell d}(t)=[\sin(t),\dots,\sin(t)]^{\top}\in\mathcal{Q}_{\ell}$ as a desired links trajectory  and  $\Pi(\tilde{q},t)=\Lambda:=\text{diag}\{\Lambda_{\ell},\Lambda_m\}$ as the position contraction metric,  where $\Lambda_{\ell}$ and $\Lambda_m$ are constant\footnote{Constructing non-constant contraction metrics is not easy in general. However, some procedures have been proposed in the literature; we refer to the interested reader on the construction of a state-dependent matrix $\Pi_{\tilde{q}_v}(\tilde{q}_v,t)$ to the works of Sanfelice\cite{sanfeliceconvergence2} and Kawano\cite{Kawano}, and references therein.} and positive definite diagonal matrices

	\subsection{Experimental setup}
\begin{figure}[h!]
		\centering		
		\includegraphics[width=7cm]{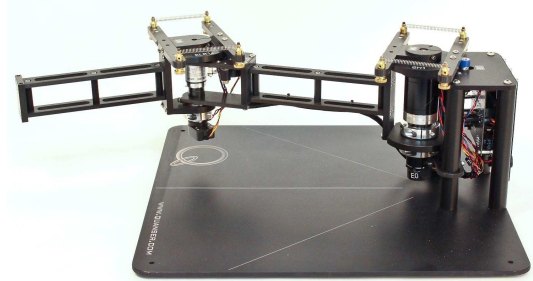}
		\caption{Quanser 2 degrees of freedom serial flexible joints robot manipulator.}
		\label{fig:quanser}		
	\end{figure}

	The experimental setup consists of a two degrees of freedom planar flexible-joints robot from Quanser \cite{quanser}; see  Figure \ref{fig:quanser}. 

	 	For the FJR in  Figure \ref{fig:quanser} we have that $n_{\ell}=n_m=2$ in  \eqref{eq:phmechanical-flexible}, and its parameters   are shown in Table \ref{table:quanserparameterslink}:
 
	\begin{table}[h!]
		\centering	
		\begin{tabular}{||c| c || c| c||c|c||} 
			\hline
			Parameter & Value & Parameter  & Value & Parameter & Value\\ [0.5ex] 
			\hline\hline
			$m_{\ell 1}$& $1.510 kg$ &   $I_{\ell 1}$ &  $0.0392kg\cdot m^2$ & $\ell_{\ell 1}$ &  $0.343m$  \\ 
			\hline
			$m_{\ell 2}$ & $0.873 kg$  &  $I_{\ell 2}$& $0.00808 kg\cdot m^2$   &  $\ell_{\ell 2}$ & $0.267m$   \\ 
			\hline
			$m_{m 1}$ & $0.23 kg$  & $r_{\ell 1}$ &  $0.159m$  & $D_{\ell}$ &  diag$\{0.8,0.55\}N\cdot s/m$ \\ 
			\hline			 			 
			$m_{m 2}$ & $0.01 kg$&  $r_{\ell 2}$ & $0.055m$  & $D_{m}$&  diag$\{  0.2, 90\}N\cdot s/m$   \\ 
			\hline 				
		\end{tabular}
		\caption{The parameter values  of Quanser FJR as shown in Figure \ref{fig:quanser}}\label{table:quanserparameterslink}								
	\end{table}

	\noindent	The links and motor inertia matrices are  
	\begin{equation}
	M_{\ell}(q_{\ell})=\begin{bmatrix}
	a_1+a_2+2b\cos(q_{\ell2}) & a_2+b\cos(q_{\ell2})\\
	a_2+b\cos(q_{\ell2}) & a_2
	\end{bmatrix} \quad\text{and}\quad M_m(q_m)=\begin{bmatrix}
	m_{m1} & 0_{n_m}\\
	0_{n_m} & m_{m2}
	\end{bmatrix},
	\end{equation}
	respectively; with  $a_1= m_{\ell 1}r_{\ell 1}^2+m_{\ell 2}\ell_{\ell 1}^2+I_{\ell 1}, 
	a_2=m_{\ell 2} r_{\ell 2}^2+I_{\ell 2}, 
	b=m_{\ell 2}\ell_{\ell 1}r_{\ell 2}$. 
	The workless forces matrix  \eqref{eq:workless-flexible}  is
	\begin{equation}
	E(x)=b \sin(q_{\ell 2})\begin{bmatrix}
	\dot{q}_{\ell 2} & - \dot{q}_{\ell 1} & 	0_{n_m} & 	0_{n_m}\\
	\left(\dot{q}_{\ell 1}+\dot{q}_{\ell 2} \right) & 	0_{n_{\ell}} & 	0_{n_m} & 	0_{n_m}\\
	0_{n_{\ell}} & 	0_{n_{\ell}} & 	0_{n_m} & 	0_{n_m}\\	
	0_{n_{\ell}} & 	0_{n_{\ell}} & 	0_{n_m} & 	0_{n_m}\\	
	\end{bmatrix}_{\dot{q}=M^{-1}(q)p},
	\end{equation}	
	whose structure's block matrices are  explicitly given by
	\begin{equation}
	S_{\ell}=b \sin(q_{\ell 2})\begin{bmatrix}
	0_{n_{\ell}} & -  \dot{q}_{\ell 1}-0.5\dot{q}_{\ell 2}\\
	\dot{q}_{\ell 1}+0.5\dot{q}_{\ell 2} &	0_{n_{\ell}}
	\end{bmatrix}, \dot{M}_{\ell} = -b \sin(q_{\ell 2}) \begin{bmatrix}
	2 \dot{q}_{\ell 2} &  \dot{q}_{\ell 2}\\
	\dot{q}_{\ell 2} & 0_{n_{\ell}}
	\end{bmatrix}, S_m =\begin{bmatrix}
	0_{n_m}& 0_{n_m}\\
	0_{n_m} & 0_{n_m}
	\end{bmatrix}, \dot{M}_m =\begin{bmatrix}
	0_{n_m}& 0_{n_m}\\
	0_{n_m} & 0_{n_m}
	\end{bmatrix}.
	\end{equation}

	\subsection{A saturated-type $(\Lambda,K_d,\phi_1(\tilde{q}_{v}))$-controller}   
	This scheme  is an example of  Corollary \ref{corollary:preserving-variational-pH-structure}  where  only the pH-like variational structure in  \eqref{eq:variationalvirtualpHsystem} is preserved in the closed-loop.  Let us introduce the following operators for given  vector $w\in\mathds{R}^p$ as
	\begin{equation}
	\text{Tanh}(w):=\begin{bmatrix}
	\text{tanh}(w_1),\\
	\vdots\\
	\text{tanh}(w_p)
	\end{bmatrix}\in\mathds{R}^p \quad\quad\text{and}\quad\quad \text{SECH}(w)=\begin{bmatrix}
	\text{sech}(w_1)& \cdots& 0\\
	\vdots&\ddots & \vdots\\
	0 & \cdots & 	\text{sech}(w_p)
	\end{bmatrix}\in\mathds{R}^{p\times p}.
	\end{equation}
	
	\subsubsection{Controller construction}		
	Since  conditions on $\Pi_q$ and $K_d$ are already given, the constructive procedure  is reduced to  finding   $\phi_{\ell}(\tilde{q}_{\ell v})$ and $\phi_m(\tilde{q}_{mv})$ such that inequalities 
	in  \eqref{eq:controllaw-positionmetricinquality} and \eqref{eq:controllaw-positionmetricinquality-motor} hold simultaneously, or equivalently a function $\phi_1(\tilde{q}_v)=[\phi_{\ell}^{\top}(\tilde{q}_{\ell v}),\phi_m^{\top}(\tilde{q}_{mv})]^{\top}$ such that 
	\begin{equation}
	-\Lambda\frac{\partial \phi_1}{\partial \tilde{q}_v}(\tilde{q}_v)-\frac{\partial \phi_1^{\top}}{\partial \tilde{q}_v}(\tilde{q}_v)\Lambda\leq-2\beta_{\tilde{q}}\Lambda.
	\label{eq:contraction-inequality-compact-experiments}
	\end{equation}

	\begin{corollary}\label{corollary:controller-tanh}
		Consider   $\phi_1(\tilde{q}_v):=\Lambda \text{Tanh}(\tilde{q}_v)$.  Then, hypotheses in Corollary \ref{corollary:preserving-variational-pH-structure} hold and inequality  \eqref{eq:contraction-inequality-compact-experiments} is satisfied with
		\begin{equation}
		\beta_{\tilde{q}}=\frac{\lambda_{\min}(\Lambda^2)\cdot\lambda_{\min}(\text{SECH}^2(\tilde{q}_v))}{\lambda_{\max}(\Lambda)},
		\end{equation}   		 	 
		where $\lambda_{\min}(\cdot)$ and $\lambda_{\max}(\cdot)$ are the minimum and maximum eigenvalue of their matrix argument, respectively.
	\end{corollary}
	Notice that despite the pH-like structure  of \eqref{eq:pHlike-virtual} is \emph{ not preserved}, the vector field  $\phi_1(\tilde{q}_v)$ is   a  conservative vector field.  Indeed,  
	\begin{equation}
	P_{v}(\tilde{q}_v)=\int_{0}^{\tilde{q}_v}\Lambda \text{Tanh}(\xi)d\xi=\sum_{k=1}^{n_{\ell}}\lambda_k\ln(\cosh(\tilde{q}_{\ell v,k}))+\sum_{k=1}^{n_m}\lambda_k\ln(\cosh(\tilde{q}_{mv,k})).
	\label{eq:Potential-variationalpreserving}
	\end{equation}	 
	This scalar function can be   interpreted as the true potential energy  when constrained to the manifold $\sigma_{v}=0_n$.\\
	
	\begin{remark}
		The range of $\text{sech}(\cdot)$ is $(0,1]$. Then, it implies that   $\phi_2(\tilde{q}_v)=\Lambda\tilde{q}_v$ also satisfies inequalities in \eqref{eq:controllaw-positionmetricinquality} and \eqref{eq:controllaw-positionmetricinquality-motor} with  
		\begin{equation}
		\beta_{\tilde{q}}=\frac{\lambda_{min}(\Lambda^2)}{\lambda_{max}(\Lambda)}.
		\end{equation}
		With $\phi_2(\tilde{q}_v)=\Lambda\tilde{q}_v$ condition \eqref{eq:pH-like-structure-preserving-condition} holds and the pH-like form \eqref{eq:pHlike-virtual} \emph{is preserved}, where the  Hamiltonian function in \eqref{eq:closed-looperror-Hamiltonianfunction}   is  
		\begin{equation}
		\tilde{H}_v(\tilde{x}_v,{x})=\frac{1}{2}\tilde{q}_{}^{\top}\Lambda \tilde{q} +\frac{1}{2} \sigma_{   }^{\top}M^{-1}(q)\sigma_{ }.
		\label{eq:closed-looperror-Hamiltonianfunction-quadratic}
		\end{equation}		
		Hence, the scheme with $\phi_2(\tilde{q}_v)$ is a structure preserving \emph{passivity-based controller} for the original FJR. This controller is   in fact  the example presented in our preliminary conference work  in Reyes-B\'aez \cite{reyes2017virtual}, and the generalization to the FJRs case of the tracking scheme for fully-actuated rigid robots developed in Reyes-B\'aez\cite{reyesIFAC2017}. 
	\end{remark}

	\subsubsection{Experimental results}
\begin{figure}[h!]
		\centering 
		\includegraphics[width=0.9 \textwidth]{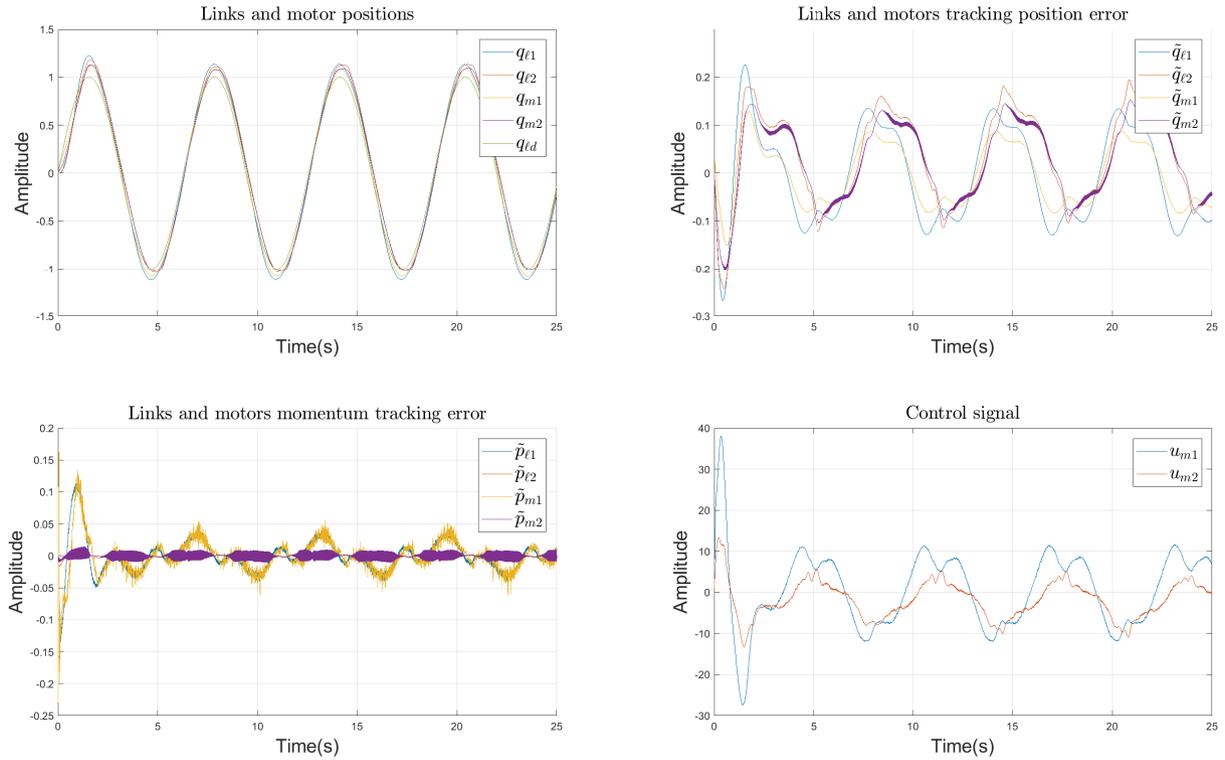}
		\caption{Closed-loop trajectories and control signal with the saturated-type $(\Lambda,K_d,\phi_1(\tilde{q}_{v}))$-controller.}
		\label{fig:controller-Tanh}	
	\end{figure}

	The experimental results of the  robot of Figure \ref{fig:quanser} in closed-loop system with this saturated-type $(\Lambda,K_d,\phi_1(\tilde{q}_{v}))$-controller  are shown in Figure \ref{fig:controller-Tanh}. The  gain matrices are $\Lambda_{\ell}=\text{diag}\{55,30\}$,  $\Lambda_{m}=\text{diag}\{70,60\}$, $K_{\ell d}=\text{diag}\{15,10\}$ and $K_{m d}=\text{diag}\{10,5\}$.\\
	On the two  upper figures, the time response of ${q}$  and $\tilde{q}_m$  is shown. On the left upper plot ${q}_{\ell}$  and $q_m$  are compared with the desired trajectory $q_{\ell d}$;  it can be seen  that links and motors positions indeed converge to $q_{\ell d}$, but only practically due to there are steady-state errors. These offsets in the state variables are attributed to the noise induced by the numerical computation of higher order derivatives.  These   can  be better observed in the upper right plot,  where the error variables are shown.  \\
	On the lower left plot of  Figure \ref{fig:controller-Tanh}, similarly, we observe that  the time response of the momentum error variables also converge practically to zero and there is noise in the signals. As said before, the main reason is that  the velocity (and hence  the momentum) are computed numerically through a filter block in Simulink which causes  some noise. \\
	Even though the family of controllers of Proposition \ref{proposition:fullstatecontroller-flexible} requires the computation of the second and third derivatives  of $q_{\ell}$ due to the definition of $p_{mr}$ in \eqref{eq:auxiliarreference-motor}, we were able to  implement controller without them by employing directly the dynamical equations in \eqref{eq:phmechanical-flexible}. In fact, the control signals are shown in the right-lower plot in Figure  \ref{fig:quanser}.

	\subsection{A v-CBC $(\Lambda,K_d,\phi_3(\cdot))$-controller via  the  matrix measure $\mu_1$} 
	By exploiting the equivalence relation between condition \eqref{eq:finslerlyapunovinequality} in the direct differential Lyapunov method of Theorem \ref{theo:lyapunovcontraction} and its counterpart for  \emph{generalized Jacobian} in 	\eqref{eq:generalizedJacobian} in terms of matrix measures, we propose an alternative  constructive procedure for   $\phi_{\ell}(\tilde{q}_{\ell v})$ and $\phi_m(\tilde{q}_{mv})$ such that conditions  \eqref{eq:controllaw-positionmetricinquality} and \eqref{eq:controllaw-positionmetricinquality-motor} are both satisfied. In this specific case, we consider the matrix measure associated to the $\|\Theta x\|_1$ norm  for a given matrices  $\Theta, A\in\mathds{R}^{p\times p}$ defined as\cite{russo2010global}
	\begin{equation}
	\mu_{1}(A):=\max_j \left(A_{jj}(\tilde{q}_v,t)+\sum_{i\neq j}|A_{ij}(\tilde{q}_v,t)|\right).
	\label{eq:matrixMeasure1}
	\end{equation}

	\subsubsection{Controller construction}
	The generalized Jacobian for $\phi_3(\tilde{q}_v)=[\phi_{\ell}^{\top}(\tilde{q}_{\ell v}),\phi_m^{\top}(\tilde{q}_{mv})]^{\top}$ in this case is 
	\begin{equation}
	\overline{J}(\tilde{q}_v,t)=\Theta\frac{\partial \phi_{3}}{\partial \tilde{q}_v}(\tilde{q}_v)\Theta^{-1}=\begin{bmatrix}
	-\frac{\partial \phi_{\ell 1}}{\partial \tilde{q}_{\ell v1}}(\tilde{q}_{\ell v})  & -\frac{\theta_1}{\theta_2}\frac{\partial \phi_{\ell 1}}{\partial \tilde{q}_{\ell v2}}(\tilde{q}_{\ell v})  & 0_{n_{\ell}}& 0_{n_{\ell}}\\
	-\frac{\theta_2}{\theta_1}\frac{\partial \phi_{\ell 2}}{\partial \tilde{q}_{\ell v1}}(\tilde{q}_{\ell v})  & -\frac{\partial \phi_{\ell 2}}{\partial \tilde{q}_{\ell v2}}(\tilde{q}_{\ell v})  & 0_{n_{\ell}}& 0_{n_{\ell}}\\	
	0_m & 0_m & -\frac{\partial \phi_{m 1}}{\partial \tilde{q}_{m v1}}(\tilde{q}_{m v}) & -\frac{\theta_3}{\theta_4}\frac{\partial \phi_{m 1}}{\partial \tilde{q}_{m v2}}(\tilde{q}_{m v})\\
	0_m & 0_m & -\frac{\theta_4}{\theta_3}\frac{\partial \phi_{m 2}}{\partial \tilde{q}_{m v1}}(\tilde{q}_{m v}) &  -\frac{\partial \phi_{m 2}}{\partial \tilde{q}_{m v2}}(\tilde{q}_{m v})	
	\end{bmatrix}, 
	\end{equation}
	where   $ \Lambda=\Theta^{\top}\Theta$   for   matrix  $\Theta=\text{diag}\{\theta_1,\theta_2,\theta_3,\theta_4\}>0_n$, and   matrix measure is explicitly given by
	\begin{equation}
	\begin{split}
	\mu_1(\overline{J})=\max\left\{
	-\frac{\partial \phi_{\ell 1}}{\partial \tilde{q}_{\ell   v1}}+\bigg|\frac{\theta_2}{\theta_1}\frac{\partial \phi_{\ell 2}}{\partial \tilde{q}_{\ell v1}}\bigg|,-\frac{\partial \phi_{\ell 2}}{\partial \tilde{q}_{\ell v2}}+\bigg|\frac{\theta_1}{\theta_2}\frac{\partial \phi_{\ell 1}}{\partial \tilde{q}_{\ell v2}}\bigg|,
	-\frac{\partial \phi_{m 1}}{\partial \tilde{q}_{m  v1}}+\bigg|\frac{\theta_4}{\theta_3}\frac{\partial \phi_{m 2}}{\partial \tilde{q}_{ml v1}}\bigg|,-\frac{\partial \phi_{m 2}}{\partial \tilde{q}_{m v2}}+\bigg|\frac{\theta_3}{\theta_4}\frac{\partial \phi_{m 1}}{\partial \tilde{q}_{m v2}}\bigg|\right\}.
	\end{split}
	\label{eq:matrixmeasure1-explicit}
	\end{equation}	
	Thus, the contractivity condition in \eqref{eq:contraction-inequality-compact-experiments} is equivalent to 
	\begin{equation}
	\mu_{1}(\overline{J}(\tilde{q}_v,t))\leq -2\beta_{\tilde{q}_v},
	\label{eq:Experiment-contraction-condition-norm1}
	\end{equation}	
	where $2\beta_{\tilde{q}_v}:=\min\{c_1^2,c_2^2, c_3^2, c_4^2\}$, with $c_1,c_2,c_3,c_4$ positive constants   satisfying the following inequalities
	\begin{equation}
	\begin{split}
	\overline{J}_{11}(\tilde{q}_v)+|\overline{J}_{21}(\tilde{q}_v)|<-c_1^2;\quad \overline{J}_{22}+|\overline{J}_{12}|<-c_2^2;\quad \overline{J}_{33}(\tilde{q}_v)+|\overline{J}_{43}(\tilde{q}_v)|<-c_3^2;\quad \overline{J}_{44}+|\overline{J}_{34}|<-c_4^2.
	\end{split}
	\label{eq:matrixmeasure1}	
	\end{equation}
	
	\begin{corollary}
		Let $\phi_3(\tilde{q}_v)$ be defined by 
		\begin{equation}
		\phi_3(\tilde{q}_v)=\begin{bmatrix}
		\phi_{\ell 1}(\tilde{q}_{\ell v})\\
		\phi_{\ell 2}(\tilde{q}_{\ell v})\\		
		\phi_{m 1}(\tilde{q}_{mv})\\		
		\phi_{m 2}(\tilde{q}_{m v})				
		\end{bmatrix}=\begin{bmatrix}
		(1+\kappa_1)\tilde{q}_{\ell v1}+\frac{\theta_2}{\theta_1}\text{tanh}(\tilde{q}_{\ell v2})\\
		\frac{\theta_1}{\theta_2}\text{tanh}(\tilde{q}_{\ell v1})+(1+\kappa_2)\tilde{q}_{\ell v2}\\
		(1+\kappa_3)\tilde{q}_{mv1}+\frac{\theta_4}{\theta_3}\text{tanh}(\tilde{q}_{mv2})\\
		\frac{\theta_3}{\theta_4}\text{tanh}(\tilde{q}_{mv1})+(1+\kappa_4)\tilde{q}_{mv2}		
		\end{bmatrix},
		\label{eq:matrixmeasure1-phi-flexible}
		\end{equation}	
		where  $\kappa_1,\kappa_2,\kappa_3,\kappa_4$ are strictly positive constants. Then, condition  \eqref{eq:Experiment-contraction-condition-norm1} is satisfied with $c_1^2=\kappa_1, c_2^2=\kappa_2,c_3^2=\kappa_3$ and $c_4^2=\kappa_4$		.
	\end{corollary}
	With this scheme neither  the structure of \eqref{eq:pHlike-virtual} nor the variational one of  \eqref{eq:variationalvirtualpHsystem}  are preserved. Nevertheless,  uniform global exponential convergence to $q_{\ell d}$ is still guarantee. Interestingly, in this scheme the convergence  rate $\beta_{\tilde{q}_v}$  does not depend on gain $\Lambda$,  which give extra  freedom in the tuning process. In particular,  when constrained to the manifold $\sigma_v=0_n$, the convergence to $q_{\ell d}$ can be accelerated by the gain $\kappa_i, i\in\{1,\dots,4\}$. 
	
	\subsubsection{Experimental results}	
\begin{figure}[h!]
		\centering 
		\includegraphics[width=0.9 \textwidth]{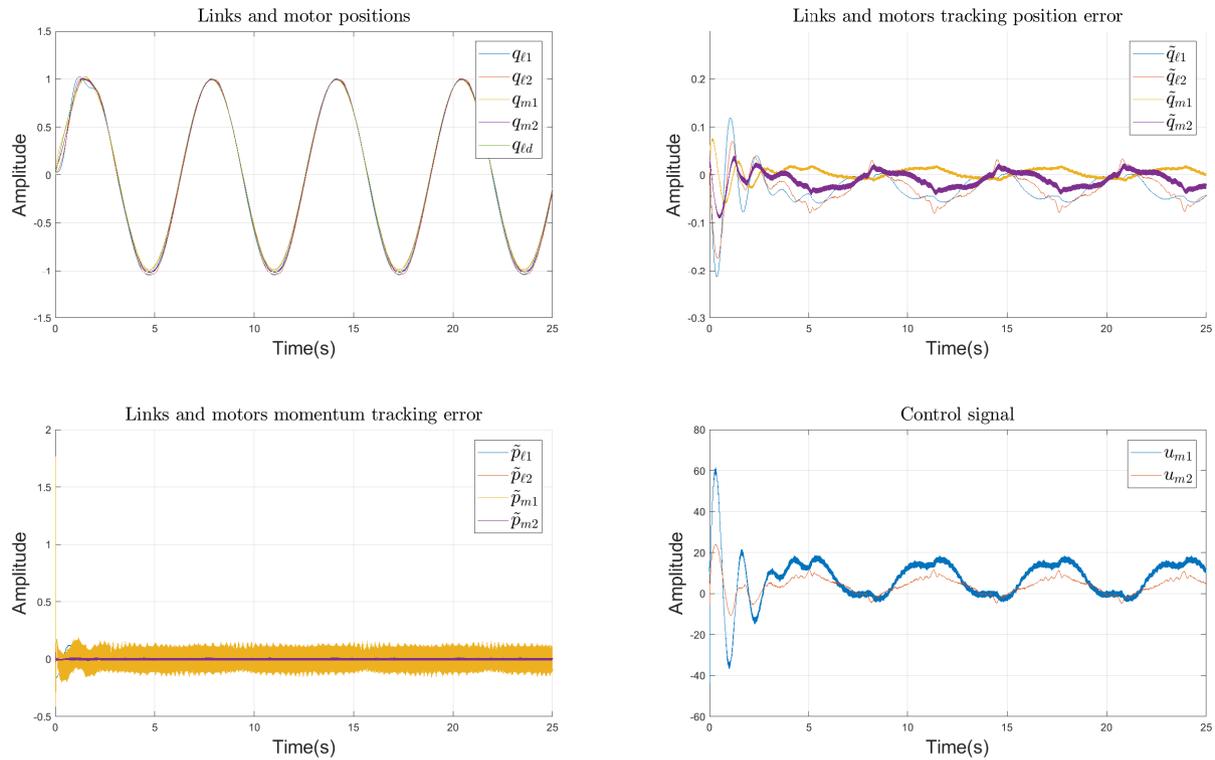}
		\caption{Closed-loop trajectories and control signal with the  $(\Lambda,K_d,\phi_1(\tilde{q}_{v}))$-controller via the  matrix measure $\mu_1$.}
		\label{fig:Kappa-control}	
	\end{figure}

	For the experiment with  this controller, we  consider the following specifications: $\kappa_1=10,\kappa_2=8$,  $\theta_1=\sqrt{\Lambda_{\ell,11}}$, $\theta_2=\sqrt{\Lambda_{\ell, 22}}$, $\theta_3=\sqrt{\Lambda_{m,11}}$ and  $\theta_4=\sqrt{\Lambda_{m, 22}}$ with the same gain matrices $\Lambda_{\ell}$, $\Lambda_m$, $K_{\ell d}$ and $K_{md}$  of the previous experiment. \\	
	The closed-loop time response is shown in Figure \ref{fig:Kappa-control}. At first stage we can observe that the performance  with respect to the previous controller  is improved; this is mainly attributed to the  gains $\kappa_i, i\in\{1,\dots,4\}$.\\	
	Indeed, on the left upper plot we can see how the links and motors positions \emph{almost} superimpose   the desired links trajectory $q_{\ell d}$.  This can be appreciated better on the upper-right plot where the  error variables are shown; we observe that we still have only \emph{practical convergence} since there is  steady-state errors, but these are considerably reduced with respect to the precious scheme as well as the overshoot in the transient time interval. We also observe some noise in the motors positions.\\ 	
	On the left lower plot we see the time response of the momentum error variables which   have considerably decreased with respect to the previous controller. In fact, as it may be expected the  overshoot during the transient time has decreased as well as the steady state momentum errors which amplitudes, excepting $\tilde{p}_{m1}$, is of the order of $10^{-2}$. Here we still have the noise problem due to the numerical computation of the momentum feedback, and in this case also the control effort of the links dynamics. \\
	On the right lower plot, we see that   the overshoot of the control signals has  increased but  steady-state signals amplitude is more less the same but with a \emph{rms} value added. This is the expected price to pay after adding an extra control gain.

	\section{Conclusions }
	
	In this work we have proposed a large family of virtual-contraction based controllers that solve the standard trajectory tracking problem of FJRs modeled as  port-Hamiltonian systems. With these controllers, global exponential convergence to a predefined reference trajectory is guaranteed. The design procedure is based on the notions of contractivity and virtual systems. 
	
	The developed family of v-CBC are PD-like controllers which have three design "parameters" that  give different structural properties to the closed-loop virtual system like  pH-like structure preserving, variational pH-like structure preserving, differential passivity, among others. These properties were used for constructing  two \emph{novel} nonlinear PD-like v-CBC schemes.  The performance of the aforementioned controllers was evaluated experimentally using  the  planar flexible-joints  robot of two degrees of freedom by from Quanser. \\

%
%

	\bibliography{bibliografia.bib}%

\begin{thebibliography}{10}

\bibitem{tomei1995tracking}
Nicosia S, Tomei P. A tracking controller for flexible joint robots using only
  link position feedback.  {\it IEEE Transactions on Automatic Control.
  }1995;40(5).

\bibitem{spongflexible}
Spong M~W. Modeling and control of elastic joint robots.  {\it Journal of
  dynamic systems, measurement, and control. }1987;109(4):310--319.

\bibitem{canudas}
Canudas {de}~Wit C, Siciliano B, Bastin G. {\it Theory of robot control}.
\newblock Springer Science \& Business Media; 2012.

\bibitem{loria1995tracking}
Loria A, Ortega R. On tracking control of rigid and flexible joints robots.
  {\it Appl. Math. Comput. Sci. }1995;5(2):101--113.

\bibitem{ortega-regulacion}
Ailon A, Ortega R. An observer-based set-point controller for robot
  manipulators with flexible joints.  {\it Systems \& Control Letters.
  }1993;21(4):329 - 335.

\bibitem{brogliato1995global}
Brogliato B, Ortega R, Lozano R. Global tracking controllers for flexible-joint
  manipulators: a comparative study.  {\it Automatica. }1995;31(7):941--956.

\bibitem{ortega2013passivity}
Ortega R, Perez J~A, Nicklasson P~J, Sira-Ramirez H. {\it Passivity-based
  control of Euler-Lagrange systems}.
\newblock Springer Science \& Business Media; 2013.

\bibitem{astolfi2003immersion}
Astolfi A, Ortega R. Immersion and invariance: A new tool for stabilization and
  adaptive control of nonlinear systems.  {\it IEEE Transactions on Automatic
  control. }2003;48(4):590--606.

\bibitem{albu2007unified}
Albu-Sch{\"a}ffer A, Ott C, Hirzinger G. A unified passivity-based control
  framework for position, torque and impedance control of flexible joint
  robots.  {\it The international journal of robotics research. }2007;26(1).

\bibitem{Sofia}
Avila-Becerril S, Lor\'a A, Panteley E. Global position-feedback tracking
  control of flexible-joint robots.  {\it Paper presented at: American Control
  Conference (ACC), 2016. }2016; Boston, MA, USA;:3008--3013.

\bibitem{pan2017adaptive}
Pan Yongping, Wang Huiming, Li~Xiang, Yu~Haoyong. Adaptive command-filtered
  backstepping control of robot arms with compliant actuators.  {\it IEEE
  Transactions on Control Systems Technology. }2017;26(3):1149--1156.

\bibitem{vanderschaft1995}
{van der}~Schaft A~J, Maschke B~M. The Hamiltonian formulation of energy
  conserving physical systems with external ports.  {\it {Archiv f\"ur
  Elektronik und {\"U}bertragungstechnik}. }1995;49.

\bibitem{Borja2014flexible}
Ortega R, Borja L~P. New results on control by interconnection and
  energy-balancing passivity-based control of port-Hamiltonian systems.  {\it
  Paper presented at: Decision and Control (CDC), IEEE 53rd Annual Conference
  on. }2014;:2346--2351.

\bibitem{castanos}
Ortega R, {van der}~Schaft A~J, Casta{\~n}os F, Astolfi A. Control by
  interconnection and standard passivity-based control of port-Hamiltonian
  systems.  {\it IEEE Transactions on Automatic Control. }2008;53.

\bibitem{chinoflexiblejoint}
Zhang Q, Xie Z, Kui S, Yang H, Minghe J, Cai H. Interconnection and damping
  assignment passivity-based control for flexible joint robot.  {\it Paper
  presented at: Intelligent Control and Automation (WCICA), 11th World Congress
  on Intelligent Control and Automation. }2014;:4242--4249.

\bibitem{escobar}
Ortega R, {van der}~Schaft A~J, Maschke B, Escobar G. Interconnection and
  damping assignment passivity-based control of port-controlled Hamiltonian
  systems.  {\it Automatica. }2002;38(4):585--596.

\bibitem{jayawardhana2006}
Jayawardhana B. {\it Tracking and Disturbance Rejection of Passive Nonlinear
  Systems}.
\newblock Ph.D. thesis, Imperial College London; 2006.

\bibitem{hilde}
Jard\'on-Kojakhmetov H, Munoz-Arias M, Scherpen J~M~A. Model reduction of a
  flexible-joint robot: a port-Hamiltonian approach.  {\it IFAC-PapersOnLine.
  }2016;49(18):832 - 837.
\newblock 10th IFAC Symposium on Nonlinear Control Systems NOLCOS.

\bibitem{reyes2017virtual}
Reyes-B\'aez R, {van der}~Schaft A~J, Jayawardhana B. Virtual Differential
  Passivity based Control for Tracking of Flexible-joints Robots.  {\it
  IFAC-PapersOnLine. }2018;51(3):169 - 174.
\newblock 6th IFAC Workshop on Lagrangian and Hamiltonian Methods for Nonlinear
  Control LHMNC.

\bibitem{reyesIFAC2017}
Reyes-B\'aez R, {van der}~Schaft A~J, Jayawardhana B. Tracking Control of
  Fully-actuated port-Hamiltonian Mechanical Systems via Sliding Manifolds and
  Contraction Analysis.  {\it IFAC-PapersOnLine. }2017;50(1):8256 - 8261.
\newblock 20th IFAC World Congress.

\bibitem{RodoAutomatica2017}
Reyes-B\'aez R, {van der}~Schaft A~J, Jayawardhana B. Virtual differential
  passivity based control for a class of mechanical systems in the
  port-Hamiltonian framework.  {\it Submitted. }2018;.

\bibitem{forni}
Forni F, Sepulchre R. A Differential Lyapunov Framework for Contraction
  Analysis.  {\it IEEE Transactions on Automatic Control. }2014;.

\bibitem{pavlov2017convergent}
Pavlov A, {van de}~Wouw Nathan. Convergent systems: nonlinear simplicity.  In:
  Springer 2017 (pp. 51--77).

\bibitem{slotinecontraction}
Lohmiller W, Slotine J~J~E. On contraction analysis for non-linear systems.
  {\it Automatica. }1998;.

\bibitem{sontag2010contractive}
Sontag E~D. Contractive systems with inputs.  In: Springer 2010 (pp. 217--228).

\bibitem{wang}
Wang W, Slotine J~J~E. On partial contraction analysis for coupled nonlinear
  oscillators.  {\it Biological cybernetics. }2005;92(1).

\bibitem{jouffroy}
Jouffroy J, Fossen T. A tutorial on incremental stability analysis using
  contraction theory.  {\it Modeling, Identification and control.
  }2010;31(3):93--106.

\bibitem{manchester2015unifying}
Manchester I~R, Tang J~Z, Slotine J~J~E. Unifying classical and
  optimization-based methods for robot tracking control with control
  contraction metrics.  {\it Paper presented at: International Symposium on
  Robotics Research (ISRR). }2015;:1--16.

\bibitem{angeli2002lyapunov}
Angeli D. A Lyapunov approach to incremental stability properties.  {\it IEEE
  Transactions on Automatic Control. }2002;47(3):410--421.

\bibitem{arjan2013differentialpassivity}
{van der}~Schaft A~J. On differential passivity.  {\it IFAC Proceedings
  Volumes. }2013;46(23):21--25.
\newblock 9th IFAC Symposium on Nonlinear Control Systems.

\bibitem{ReyesBaez-PhDthesis}
{Reyes B{\'a}ez} Rodolfo. Virtual contraction and passivity based control of
  nonlinear mechanical systems: trajectory tracking and group coordination.
\newblock PhD thesisUniversity of Groningen2019.

\bibitem{khalil1996noninear}
Khalil H~K. Noninear systems.  {\it Prentice-Hall, New Jersey. }1996;2(5):5--1.

\bibitem{crouch}
Crouch PE, {van der}~Schaft AJ. Variational and Hamiltonian Control Systems.
  {\it Springer-Verlag. }1987;.

\bibitem{sanfeliceconvergence2}
Sanfelice R~G, Praly L. Convergence of nonlinear observers on $\mathds{R}^n$
  with a Riemannian metric (part I).  {\it IEEE Transactions on Automatic
  Control. }2015;.

\bibitem{coogan2017contractive}
Coogan S. A Contractive Approach to Separable Lyapunov Functions for Monotone
  Systems.  {\it arXiv preprint arXiv:1704.04218. }2017;.

\bibitem{russo2010global}
Russo G, Di~Bernardo M, Sontag E~D. Global entrainment of transcriptional
  systems to periodic inputs.  {\it PLoS computational biology.
  }2010;6(4):e1000739.

\bibitem{forni2013differentially}
Forni F, Sepulchre R. On differentially dissipative dynamical systems.  {\it
  IFAC Proceedings Volumes. }2013;46(23):15 - 20.
\newblock 9th IFAC Symposium on Nonlinear Control Systems.

\bibitem{forni2013differential}
Forni F, Sepulchre R, {van der}~Schaft A~J. On differential passivity of
  physical systems.  In: :6580--6585IEEE; 2013.

\bibitem{Arimoto84}
Arimoto S, Miyazaki F. Stabilidty and robustness of PID feedback control for
  robot manipulators of sensory capability.  {\it Robotics Research, The 1st
  Symp., by M Brady \& R.P. Paul, Eds., MIT Press, Cabridge Massachusetts.
  }1984;.

\bibitem{ott2008passivity}
Ott C, Albu-Schaffer A, Kugi A, Hirzinger G. On the passivity-based impedance
  control of flexible joint robots.  {\it IEEE Transactions on Robotics.
  }2008;24(2):416--429.

\bibitem{pHbook}
{van der}~Schaft A~J, Jeltsema D. Port-Hamiltonian systems theory: An
  introductory overview.  {\it Foundations and Trends in Systems and Control.
  }2014;1.

\bibitem{pavlov2006}
Pavlov A, Marconi L. Incremental passivity and output regulation.  {\it Paper
  presented at: IEEE Conference on Decision and Control. }2006;.

\bibitem{Kawano}
Kawano Y, Ohtsuka T. Nonlinear Eigenvalue Approach to Differential Riccati
  Equations for Contraction Analysis.  {\it IEEE Transactions on Automatic
  Control. }2017;62(12):6497-6504.

\bibitem{quanser}
Quanser Consulting Inc. 2-DOF serial flexible link robot, Reference Manual,
  Doc. No. 763, Rev. 1, 2008.

\end{thebibliography}

\end{document}